\documentclass[twocolumn,showpacs,superscriptaddress,preprintnumbers,amsmath,amssymb,aps,prx]{revtex4-1}
\usepackage{graphicx}
\usepackage{mathptmx}
\usepackage{verbatim}
\usepackage{graphicx}
\usepackage{amsmath}
\usepackage{amssymb}
\usepackage{amsfonts}
\usepackage{mathrsfs}
\usepackage{amsmath} 
\usepackage{color}
\usepackage{graphicx}
\usepackage{bm} 
\usepackage{amssymb}
\usepackage{xspace}
\usepackage{hyperref}
\usepackage{multirow}
\usepackage{color}
\usepackage{float}
\usepackage{bbold}
\usepackage{tabu}
\usepackage{textcomp}
\usepackage{dcolumn}
\usepackage{appendix} 

\newcolumntype{L}[1]{>{\raggedright\arraybackslash}m{#1}}
\newcolumntype{C}[1]{>{\centering\arraybackslash}m{#1}}
\newcolumntype{R}[1]{>{\raggedleft\arraybackslash}m{#1}}
\newcolumntype{N}{@{}m{0pt}@{}}
\begin{document}

\title{Topological flat bands without magic angles in massive twisted bilayer graphenes}

\author{Srivani Javvaji}
\affiliation{Department of Physics, Ningbo University, Zhejiang 315211, P. R. China}
\affiliation{Department of Physics, University of Seoul, Seoul 02504, Korea}
\author{Jinhua Sun}
\email{sunjinhua@nbu.edu.cn}
\affiliation{Department of Physics, Ningbo University, Zhejiang 315211, P. R. China}
\affiliation{Department of Physics, University of Seoul, Seoul 02504, Korea}
\author{Jeil Jung}
\email{jeiljung@uos.ac.kr}
\affiliation{Department of Physics, University of Seoul, Seoul 02504, Korea}

\begin{abstract}
Twisted bilayer graphene (TBG) hosts nearly flat bands with narrow bandwidths of a few meV at certain {\em magic} twist angles.
Here we show that in twisted gapped Dirac material bilayers, or massive twisted bilayer graphenes (MTBG),
isolated nearly flat bands below a threshold bandwidth $W_c$ are expected for continuous small twist angles up to a critical $\theta_c$ 
depending on the flatness of the original bands and the interlayer coupling strength. 
Narrow bandwidths of $W \lesssim $30~meV are expected for $\theta \lesssim 3^{\circ} $ 
for twisted Dirac materials with intrinsic gaps of $\sim 2$~eV 
that finds realization in monolayers of gapped transition metal dichalcogenides (TMDC), silicon carbide (SiC) among others, 
and even narrower bandwidths in hexagonal boron nitride (BN) whose gaps are $\sim 5$~eV,
while twisted graphene systems with smaller gaps of a few tens of meV, e.g. due to alignment with hexagonal boron nitride, 
show vestiges of the magic angles behavior in the bandwidth evolution. 
The phase diagram of finite valley Chern numbers of the isolated moire bands
expands with increasing difference between the sublattice selective interlayer tunneling parameters. 
The valley contrasting circular dichroism for interband optical transitions is constructive near $0^{\circ}$ and destructive near $60^{\circ}$ alignments and
can be tuned through electric field and gate driven polarization of the mini-valleys. 
Combining massive Dirac materials with various intrinsic gaps, Fermi velocities, interlayer tunneling strengths 
suggests optimistic prospects of increasing $\theta_c$ and achieving correlated states with large $U/W$ 
effective interaction versus bandwidth ratios.
\end{abstract}
\maketitle

\section{Introduction}
The field of atomically thin two dimensional (2D) materials has seen an enormous progress over the last 
fifteen years since the successful experiments identifying the massless Dirac cones in graphene~\cite{geim,geim1,pkim}.
In recent years the research directions have shifted towards tailoring the electronic structure of 2D material
heterojunctions by selective material control and stacking arrangement~\cite{2dheterojunctionPKim,2dheterojunctionTMD1,2dheterojunctionTMD2,2dheterojunctionTMD3}.
New families of hexagonal 2D materials beyond graphene including group IV layered materials 
~\cite{FalkoSilicene,silicene,siliceneDFT,germanene1,germanene2,tinene},
semiconducting and metallic transition metal dichalcogenides (TMDC)~\cite{tmdc1,tmdc2,tmdc3,tmdc4} expand the list of the so called Dirac
materials whose relevant band structure near the Fermi level is located at the $K$ and $K'$ Dirac points at the Brillouin zone corners.
Modifications in the stacking arrangements are expected to introduce significant changes in the electronic structures of 2D materials assemblies.
One prototypical example is the twisted bilayer graphene~\cite{deheer1,deheer2,deheer3,deheer4,deheer5,ugeda,ohta,lopes,lopes1,shallcross,shallcross1,
shallcross2,shallcross3,bistritzer,koshino,koshino1,jung,sanjose,sanjose1,sanjose2,Bistritzerprb,wangzf,schmidt, stephen2018,koshino2,vafek,vishwanath,vishwanath2}
where the relative rotation of one of the layers gives rise to new features
in the electronic structure such as van Hove singularities and secondary Dirac cones~\cite{luican,andrei}
that can be understood based on the associated moire bands~\cite{bistritzer,lopes,jung,dillonwong,abhay,nadj}.
Observations of superconductivity~\cite{super1,super2,super3} and correlated phases~\cite{mott1,mott2,mott3} 
in twisted bilayer graphene have sparked a renewed interest in searching the properties of twisted bilayer graphene 
and flat bands in other types of twisted 2D materials.
New proposals for flat bands in hybrid heterostructures include transition metal dichalcogenides~\cite{fengcheng,mitmanish,naik2019,feenstra},
hexagonal boron nitride bilayers~\cite{bnbn, wei2019}, 
van der Waals patterned dielectric superlattices~\cite{song2019}, 
multilayer graphene systems such as twisted bi-bilayer graphene~\cite{tbbg,senthil,daixi}, 
and rhombohedral trilayer graphene on hexagonal boron nitride where gate tunable 
correlated and superconducting phases have been observed experimentally~\cite{guorui1,guorui2}. 
When the Dirac material layers are gapped by applying vertical electric fields~\cite{chittari2019,tbbg,tbbg1,tbbg2,tbbg3,tbbg4,tbbg5}
or when the gaps are already open like in BN/BN bilayers~\cite{bnbn},
the bandwidths remained reduced continuously for small twist angles
rather than only at a discrete set of magic angles like in twisted bilayer graphene~\cite{bistritzer}. 
In this work we investigate the behavior of the low energy bands bandwidth for different massive twisted bilayer graphene (MTBG) Hamiltonians 
and find that nearly flat bands below a threshold bandwidth $W_c$ can be achieved for a continuous range of twist angles 
up to a critical angle $\theta_c$ that increases with the band gap size and with the reduction of the Fermi velocity. 
Our study indicates that low energy nearly flat bands can be expected in the vicinity of band edges 
for small enough twist angles in a variety of gapped materials including semiconducting TMDC, silicon carbide (SiC), and hexagonal boron nitride (BN) and other 2D materials twisted 
bilayers when they can be appropriately modeled by gapped Dirac Hamiltonians. 
This flexibility in the twist angle choice should enormously facilitate the preparation of twistronics devices to tailor the flat bands. 
Furthermore, we calculated the associated topological valley Chern numbers phase diagram that can lead to 
spontaneous quantum Hall phases in the absence of magnetic fields~\cite{haldane1988,kanemele2005,nandkishore2010,jung2011,zhang2011} 
when the degeneracy of these bands are lifted under the effects of spin or valley polarising time reversal symmetry 
breaking perturbations~\cite{chittari2019,senthil,fengwang2019},
and whose signatures have been measured in recent experiments~\cite{mott3,andreayoung2019,fengwang2019}.
The interband transition and circular dichroism in twisted gapped Dirac materials is studied
as a function of twist angle, electric fields and stacking alignment between the layers to illustrate the tunability
of the optical properties in the system. 

\begin{figure*}[htb!]
	\begin{center}
		\includegraphics[width=17cm]{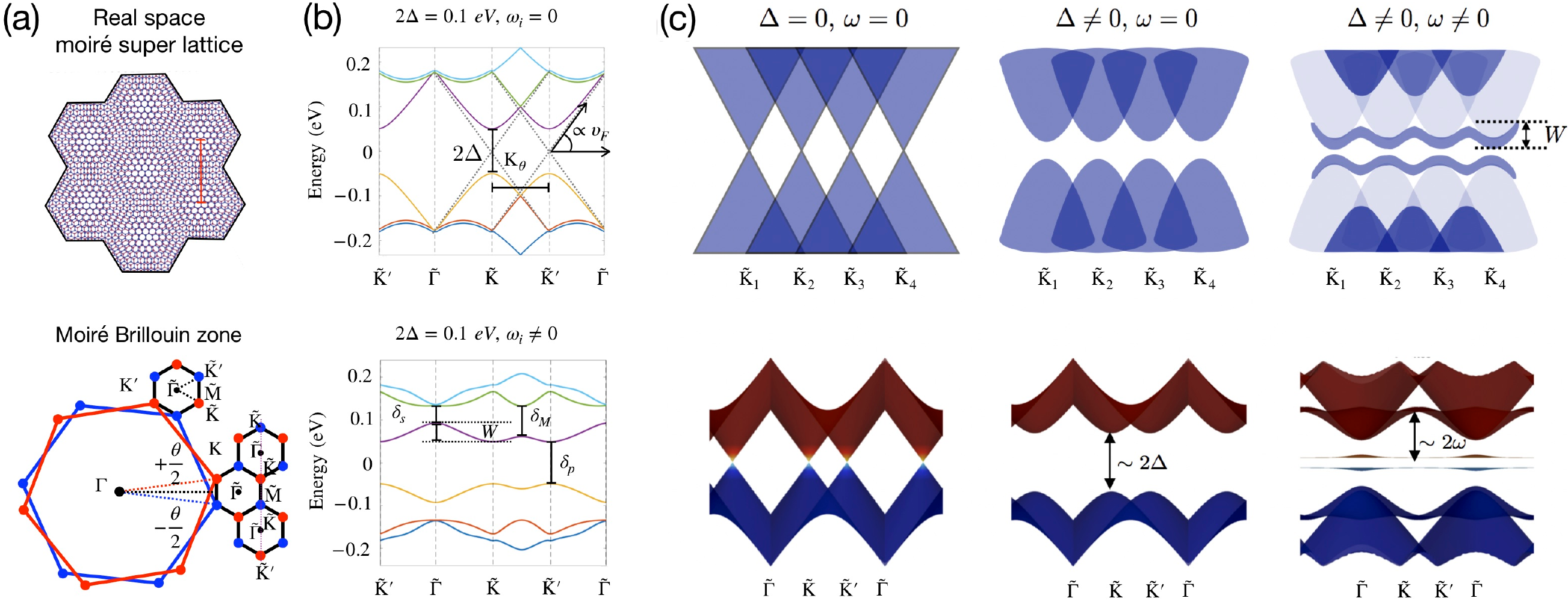}		
		\caption{(Color online) 
		Interband coherence leading to band isolation and flattening between gapped Dirac cones in the moire Brillouin zone (mBZ).
		(a) Real-space moire pattern and moire Brillouin zone showing the followed $k$-path at the mBZ of the $K$ macrovalley. 
		(b) Hyperbolic band dispersion $E(k) = (\Delta^2 + (\upsilon_{\rm F} p )^2)^{1/2}$ that is parabolic at small $p$ and linear at large 
		momenta due to an intralayer band gap $2 \Delta$.
		The rotation between the layers separates the bands in momentum space by $K_{\theta} = K \theta$ proportional to the twist angle. 
		Flatter bands are expected for smaller Fermi velocities $\upsilon_{\rm F}$ and large $\Delta$. 
		A system with finite interlayer coupling $\omega_1 = \omega_3 = 0.01$~eV and $\omega_2=0.05$~eV 
		illustrates the opening of a secondary gap $\delta_s$ near $\tilde{\Gamma}$, 
		and avoided crossings $\delta_{\rm M}$ near $\tilde{M}$, and suppressions in the bandwidth $W$. 
		(c) Schematic representation of multiple rotated Dirac cones, gapped Dirac cones, and the generation of isolated flatbands in the presence of interlayer tunneling in the top row, and side view of actual band structure surface plots in the bottom row.
		}
		\label{figure1}
	\end{center}
\end{figure*}
%
%

%
%
The manuscript is organized as follows. 
We present in Sec.~\ref{hamiltonianmodel} the theoretical details of the Hamiltonian model
used in Sec.~\ref{bandwidths} to analyze the bandwidth dependence as a function of Hamiltonian parameters.
In Sec.~\ref{analyticalsec} we discuss the analytical solutions of the eigenvalues at the symmetry points of the moire Brillouin zone. 
Then in Sec.~\ref{chernnumbers} we discuss the valley Chern number phase diagram,
and in Sec.~\ref{selectionrules} the circular dichroism of the interband transition oscillator strengths
before we close the manuscript in Sec.~\ref{summarysec} with the summary and discussions.

\section{Twisted Gapped Dirac Hamiltonian model}
\label{hamiltonianmodel}
The twisted gapped Dirac materials systems we consider are described based on 
the conventional assumption of the moire bands theory that the interlayer tunneling strength varies slowly with 
respect to the 2D position on an atomic scale~\cite{bistritzer,falkominibands,lopes,jung}. 
Our continuum Hamiltonian shares the same periodicity of the moire pattern and we employ
the Bloch's theorem to formulate the effective low energy electronic structure
in the moire Brillouin zone it defines, see Fig.~\ref{figure1}(a).
The moire bands theory used in graphene on graphene (G/G) 
and subsequent extension for more general material combinations including graphene on 
hexagonal boron nitride (G/BN) have been discussed in Ref.~\cite{jung},
where a recipe is proposed to inform the model Hamiltonian parameters from first principles calculations 
from the crystal Hamiltonian that goes beyond the two center approximation for the interatomic interactions.
%
%
The associated continuum models are generalized to include additional diagonal and off-diagonal moire 
pattern terms to the Bistritzer-MacDonald model resulting in:
\begin{eqnarray}
H_{\rm tMBG} (\bm k) = \begin{pmatrix} h^{t}_{\theta/2}({\bm k})  & T({\bm r})  \\ T^{\dag}({\bm r})  &  h^{ b}_{-\theta/2}({\bm k})  \end{pmatrix}
\label{Eq:Htbg}
\end{eqnarray}
where the bottom and top layers are twisted symmetrically in opposite senses to preserve the 
orientation of the moire pattern and therefore of the moire Brillouin zone (mBZ)
\begin{eqnarray}
h^{l}_{\theta}({\bm k})  = (\upsilon^{ l}_{\rm F} \hat{R}_{-\theta}{\bm p} + {\bm A}^{ l}({\bm r}) ) \cdot \sigma_{xy} 
+ V^{ l}({\bm r}) \mathbb{1} + \Delta^{ l}({\bm r}) \sigma_z 
\end{eqnarray}
where ${\bm p} = {\bm k} - {\bm K}$ is centered at the ${\bm K}$ point of each rotating layer, 
the index $l$ distinguishes which layer.
The triangular moire patterns for the scalar and vector potentials are given by
\begin{eqnarray}
V({\bm r}) &=& V_{0} + 2 V_{1} \, {\rm Re}\left[ e^{i \phi_V} f({\bm r}) \right], \\
\Delta ({\bm r}) &=& \Delta_{0} + 2 \Delta_{1} \, {\rm Re}\left[ e^{i \phi_{\Delta}} f({\bm r}) \right], \\
{\bm A}({\bm r}) &=&  A_{0} + 2 A_{1} \hat{z}\times\nabla {\rm Re}\left[ e^{i \phi_{\bm A}} f({\bm r}) \right], 
\end{eqnarray}
where 
$f({\bm r}) = \sum_{m=1}^6 e^{i {\bm G}_m \cdot {\bm r}} (1 + (-1)^m )/2$ that respects the triangular symmetry
is defined in terms of ${\bm G}_m = \hat{R}_{2\pi(m-1)/3} {\bm G}_1$ with $m=1, \hdots, 6$ the six first shell moire reciprocal 
lattice vectors generated rotating successively ${\bm G}_1 \simeq (0, 4\pi \theta / \sqrt{3} a)$ by an angle of $2\pi/3$~\cite{jung,laksono2017}
is equivalent to a sum of three cosines with alternating first shell G-vectors~\cite{fengcheng2019}
\begin{eqnarray}
M({\bm r}) &=& 2 C \, {\rm Re}\left[ e^{i \phi} f({\bm r}) \right]  =  2 C \sum_{m=2,4,6} \cos( {\bm G}_m {\bm r} +  \phi).
\end{eqnarray}
The operator $\hat{R}_{\theta}$ accounts for the rotation of the electrons in momentum space
through small phase differences between the top and bottom layers and is responsible for the 
chiral circular dichroism observed in TBG for twist angles in opposite senses~\cite{jiwoongpark2016,breydichroism}. 
For the analysis presented in this work we neglect these small additional phases because 
their effect is generally small and we can obtain electron-hole symmetric bands
some simple interlayer tunneling values.
We use the conventional form of the interlayer tunneling in the small angle approximation 
distinguishing the different sublattice resolved tunneling terms~\cite{bistritzer,jung}
\begin{eqnarray}
T({\bm r}) = \sum_{j} e^{ -i {\bm Q}_j {\bm r}} T^{j}_{s, s'},   \label{interlayercoupling}
\end{eqnarray}
where the three ${\bm Q}_j$ vectors ${\bm Q}_0 = K \theta (0, -1) $ and 
${\bm Q}_{\pm} = K \theta (\pm \sqrt{3}/2, 1/2)$ are proportional to twist angle $\theta$ and 
$K= 4 \pi / 3 a$ is the Brillouin zone corner length of the Dirac material of lattice constant $a$
where $a=2.46~\AA$ for graphene and $a=3.51~\AA$ for MoS$_2$. 
The interlayer coupling matrices between the two rotated adjacent layers are generically given by
\begin{equation}
T^0 =   \begin{pmatrix} \omega_1  & \omega_2  \\ \omega_2^*  &  \omega_3 \end{pmatrix},   \,\, \,\,   
T^{\pm} =  \begin{pmatrix} \omega_1  & \omega_2 e^{\mp i 2\pi/3}  \\ \omega_2^* e^{\pm i 2\pi/3}  &  \omega_3 \end{pmatrix}
\label{Eq:Tmatrix}
\end{equation}
where we  distinguish three different $\omega_1$, $\omega_2$ and $\omega_3$ 
interlayer sublattice tunneling obtained averaging over all possible stacking configurations
and assume for simplicity that they are real values. 
The convention taken here for the $T^{j}$ matrices assume an initial AA stacking configuration $\tau = (0,0)$~\cite{jung}
and differs by a phase factor with respect to the initial AB stacking $\tau = (0,a/\sqrt{3})$~\cite{bistritzer}.
We use a similar naming convention for the AA, AB and BA local stacking arrangements of the unit cell atoms
also for the 60$^{\circ}$ or equivalently 180$^{\circ}$ degrees alignment of the layers 
where the two sublattice atoms of the top layer are switched regardless of choice for the rotation center.

The moire band Hamiltonians were modeled from existing parameters in the literature for 
G/G~\cite{chittari2019,tbbg}, TMDC/TMDC~\cite{valleytmd,fengcheng2019}, 
and informed from first-principles calculation for SiC/SiC and BN/BN for this work.
The parameter fitting procedure follows closely those of Ref.~\cite{jung}
while the details for SiC/SiC are presented in appendix~\ref{fitting}.

\begin{table}
	\centering
	\caption[]{Lattice parameters, inter-site tunneling energies and band gaps of twisted gapped Dirac materials bilayer systems. 
	The Hamiltonian parameters for G/G systems are from Refs.~[\onlinecite{chittari2019}] for rigid 
	and Ref.~[\onlinecite{tbbg}] for out of plane relaxed geometries.
	Most TMDC parameters are taken from Ref.~[\onlinecite{valleytmd}] except for WSe$_2$ parameters from Ref.~[\onlinecite{fengcheng2019}], 
	and the BN and SiC parameters have been calculated from the LDA-DFT band structures for this work. 
	The intralayer moire patterns have been considered when modeling bilayers of BN, SiC, and WSe$_2$, 
	and all energies are given in eV.
	}
	\begin{tabular}{|c|c|c|c|c|c|c|} \hline \hline
	&\multicolumn{6}{c|}{ $\,\,$ Monolayer parameters and interlayer tunneling $\,$}\\ \hline
		{Bilayer} & {a (\AA)} & {$\left| t_o \right|$  } & \multicolumn{3}{c|}{$\omega$} & {$2 \Delta$ } \\ \cline{4-6}
		&  &  & $\omega_1$ & $\omega_2$ & $\omega_3$ &  \\ \hline
		G/G  (rigid)      &  2.46 &  2.6  &  \multicolumn{3}{c|}{0.098} & - \\ \hline
		G/G (relaxed)  &  2.46 &  3.1  & 0.098  &  0.12  & 0.098 & - \\ \hline
		MoS$_2$ /WS$_2$ & 3.195 & 1.1/1.37 & \multicolumn{3}{c|}{0.01} & 1.66/1.79 \\ \hline
		MoS$_2$ /MoS$_2$ & 3.193& 1.1 & \multicolumn{3}{c|}{0.01} & 1.66 \\ \hline
		WS$_2$ /WS$_2$ & 3.197 & 1.37 & \multicolumn{3}{c|}{0.01} & 1.79 \\ \hline		
		BN/BN (0$^\circ$)  & 2.48 & 2.5 &  0.178 & 0.147  &  0.078 & 4.58 \\ \hline 
		BN/BN (60$^\circ$)  & 2.48 & 2.5 & 0.208  &  0.148& 0.078  & 4.59 \\ \hline 
		SiC/SiC  & 3.06 & 1.7 & 0.165 & 0.413& 0.063 & 2.383 \\ \hline 
		WSe$_2$ /WSe$_2$ &  3.32 & 1.261 &0.0011 &0.0&0.0097& 1.2 \\ \hline
	\end{tabular}
\begin{tabular}{|c|c|c|c|c|c|c|} \hline 
&\multicolumn{6}{c|}{ Intralayer moire patterns }\\ \hline
{Coefficients}& A-A  & B-B &A$^\prime$-A$^\prime$& B$^\prime$-B$^\prime$& A-B&A$^\prime$-B$^\prime$ \\ \hline
&\multicolumn{6}{c|}{{BN/BN (0$^\circ$)}} \\ \hline
{$C_{0 ii}$}& 2.333 & $-$2.247 &2.333  & $-$2.247 &- &-\\ \hline
{$C_{ii}$}&0.0180 &0.0098 &0.0180  &0.0098 &0.0044 &0.0044 \\ \hline
{$\phi_{ii}$ }& $-$67.9 & $-$77.9 &67.9 &77.9 & $-$145 &145 \\ \hline
&\multicolumn{6}{c|}{{BN/BN (60$^\circ$)}} \\ \hline
{$C_{0\,ii}$}& 2.368& $-$ 2.222& $-$2.222  &2.368 &- & -\\ \hline
{$C_{ii}$}&0.0027 & 0.0026&0.0026  &0.0027 &0.0059 & 0.0059\\ \hline
{$\phi_{ii}$}& $-$85.8& 72.91 & 72.707 & $-$85.8 &120.12 & 120.12 \\ \hline
&\multicolumn{6}{c|}{{SiC/SiC}} \\ \hline
{$C_{0\,ii}$}&1.319 & $-$1.064 &1.319  & $-$1.064 & -&- \\ \hline
{$C_{ii}$}&0.0426 &$-$0.0085 & 0.0426 & $-$0.0085 &0.0&0.0 \\ \hline
{$\phi_{ii}$}&42.71 & $-$26.56 & $-$42.71 &26.56 &0.0& 0.0\\ \hline
&\multicolumn{6}{c|}{{WSe$_2$/WSe$_2$}} \\ \hline
{$C_{0\,ii}$}& 1.2& $0.0$ &  1.2& $0.0$ & -&- \\ \hline
{$C_{ii}$}& 0.0068&$0.0089$ & 0.0068 & $0.0089$ &0.0&0.0 \\ \hline
{$\phi_{ii}$}& $-$89.7& $-91.0$ & $89.7$ & 91.0&0.0& 0.0\\ \hline
\end{tabular}
\label{table1}
\end{table}

\begin{figure}[tb!]
	\begin{center}
		\includegraphics[width=8.5cm]{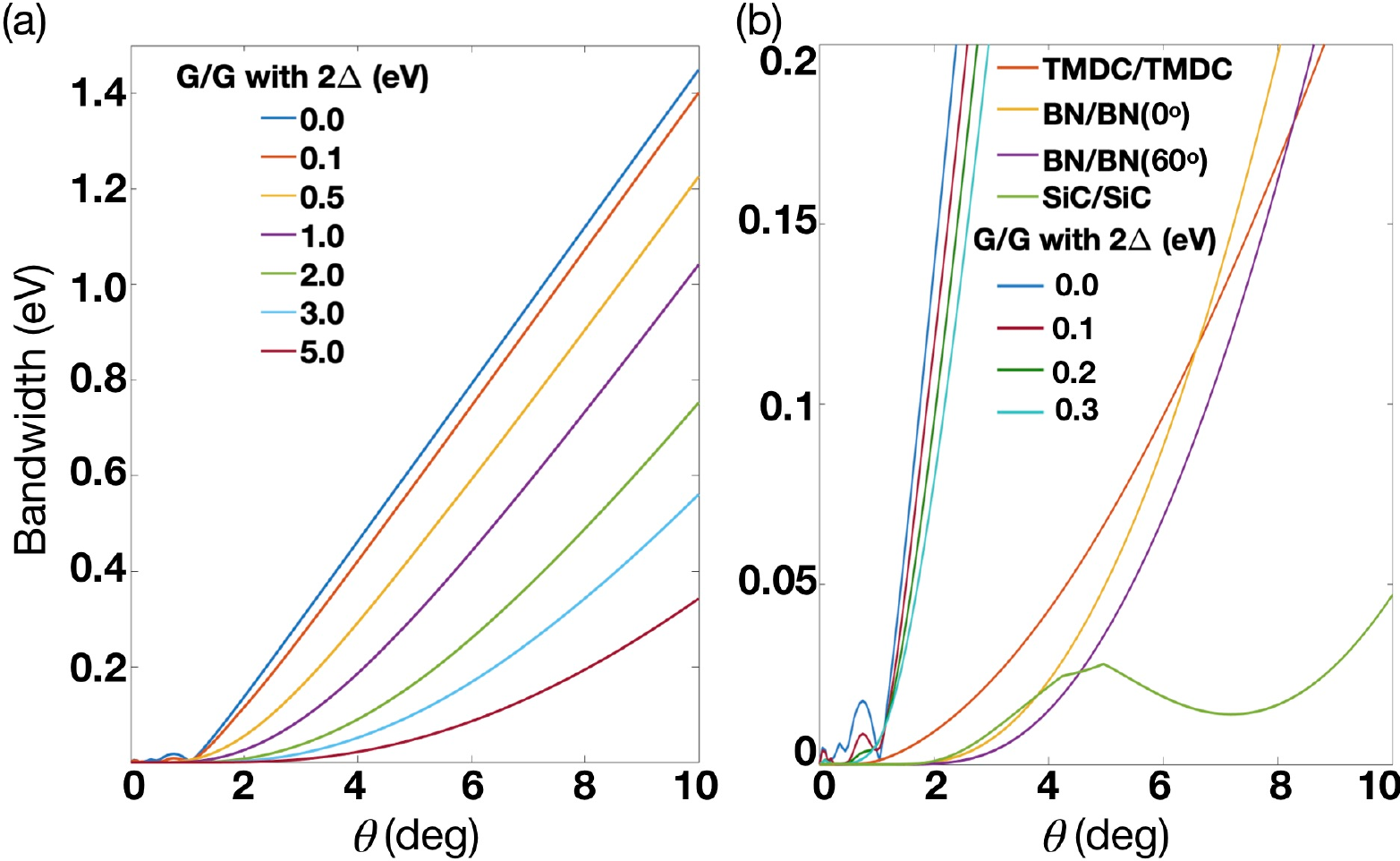}
		\caption{(Color online) 
		Bandwidth of the low energy valence flat bands as a function of twist angle in twisted gapped Dirac material
		bilayers as a function of the primary band gap magnitude for (a) massive twisted bilayer graphene, and (b) using parameters 
		of other twisted gapped Dirac material bilayers listed in Table~\ref{table1}. The material parameters for TMDC/TMDC corresponds to MoS$_2$/WS$_2$ heterobilayers.
		Traces of the magic angle bandwidth dips of gapless twisted bilayer graphene gradually disappear as the intralayer gap magnitudes become larger
		and they are almost inexistent already for gaps of the order of $\sim 0.2$~eV.
		The bands are electron-hole symmetric for equal interlayer tunneling $\omega_i= \omega$ and 
		in the absence of intralayer moire patterns and if we neglect the twist angle dependent phases.
		We observe that the bandwidths remain flat for a greater range of twist angles when the intrinsic band gaps are large.
		The non-monotonic behavior of the SiC/SiC valence band $W$ and lack of electron-hole symmetry (not shown)
		can be traced to the unequal $\omega_i$ values for the interlayer tunneling parameters.
		}
		\label{figure2}
	\end{center}
\end{figure}
\section{Bandwidth phase diagram in twisted gapped Dirac materials}
\label{bandwidths}
Here we show that the low energy moire bands in twisted gapped Dirac materials have narrow 
bandwidths for a continuous range of twist angles in contrast to the discrete set of magic angles in twisted bilayer graphene.
Specifically we show that the bandwidth of the low energy moire bands remain extremely narrow 
below a threshold value $W_c$ up to critical twist angle of $\theta_c$ 
which scales almost linearly with the intrinsic band gap $2\Delta$ and interlayer tunneling $\omega_i=\omega$, 
and is inversely proportional to the Fermi velocity $\upsilon_{\rm F}$. 
Hints of this behavior were observed in ABC-trilayer graphene on BN under a perpendicular electric field~\cite{chittari2019}, 
in the evolution of bandwidth in twisted BN/BN systems~\cite{bnbn,wei2019},
and more recently in twisted double bilayer graphene subject to electric fields~\cite{tbbg1,tbbg2,tbbg3,tbbg4,tbbg5},
and twisted transition metal dichalcogenides~\cite{fengcheng,mitmanish,naik2019,feenstra}.
%
%
To understand the behavior of the moire bands in our MTBG systems it is useful to review the 
behavior of the discrete set of magic angles in the Bistritzer-MacDonald model of twisted bilayer graphene 
where the band structure scaling parameter $\alpha = {\omega}/ ({\theta} {\upsilon_{\rm F} K})$ relating the interlayer coupling 
strength $\omega$ with the twist angle $\theta$ and the Fermi velocity $\upsilon_{\rm F}$ was used to identify the angles where the Fermi
velocity vanishes at the Dirac point~\cite{bistritzer}.
This scaling parameter $\alpha$ summarizes the relationship between the system parameters indicating that flat bands can be 
achieved more easily for greater interlayer coupling $\omega$, 
and for smaller Fermi velocities $\upsilon_{\rm F}$ and twist angles $\theta$. 
Fig.~\ref{figure1}(b) illustrates how these parameters defining $\alpha$ can affect the band structures. 
For instance, finite interlayer tunneling terms $\omega_i \neq 0$ introduce coherence between the moire bands opening a secondary $\delta_s$ 
gap at $\Gamma$ which together with the primary $\delta_p$ gap at $\tilde{K}$ allows the formation of isolated Chern bands, see Fig.~\ref{figure1}(c).
The bandwidth $W$ can be defined as the energy difference of the band energy at $\tilde{\Gamma}$ and the 
band edge at $\tilde{K}$ or $\tilde{K}'$,
and we can define $\delta_M$ as the avoided gap at the $\tilde{M}$ point located between $\tilde{K}$ and $\tilde{K}'$.
%
%

The linear relationship between $\omega$ and $\theta$ suggested by the structure of 
$\alpha$ for the first magic angle was confirmed by explicit bandwidth calculations
that numerically satisfy the $\theta_{m} = C_{m} \omega / \upsilon_{\rm F}$ relationship for 
the $m^{th}$ magic angle both in twisted bilayer graphene~\cite{chittari2019} and for the first 
magic angle corresponding to $m=1$ in the minimal model of twisted bi-bilayer graphene~\cite{tbbg} 
confirming that the magic twist angles should increase together with interlayer coupling strength 
as expected from this scaling relation~\cite{chittari2019,stephen2018}.
%
%
%
%
These magic angles were defined as the $\theta$ values that give rise to bandwidth local minima
rather than the angles where the effective Fermi velocity vanishes at the $K$-points of the mBZ~\cite{bistritzer}
since this definition would become ambiguous in twisted bi-bilayers or gapped Dirac materials 
whose band edges already have zero Fermi velocity. 
%
%

The bandwidths generally increase with twist angles and decrease with the band gap as shown in Fig.~\ref{figure2}, 
where we have represented the evolution of the bandwidth $W$ with twist angle of gapped graphene with a finite $\Delta$ mass term.
Similar evolutions of $W$ are shown also for other gapped Dirac materials that we modeled through 
other system parameters listed in Table~\ref{table1} that aims to capture the behavior of the band edges near the $K$-points, or the (macro)valleys of single layers,
of a variety of materials comprising twisted bilayers of gapped G, TMDCs including MoS$_2$ and WS$_2$, SiC, BN. 
\begin{figure}[tb!]
	\begin{center}
		\includegraphics[width=8.5cm]{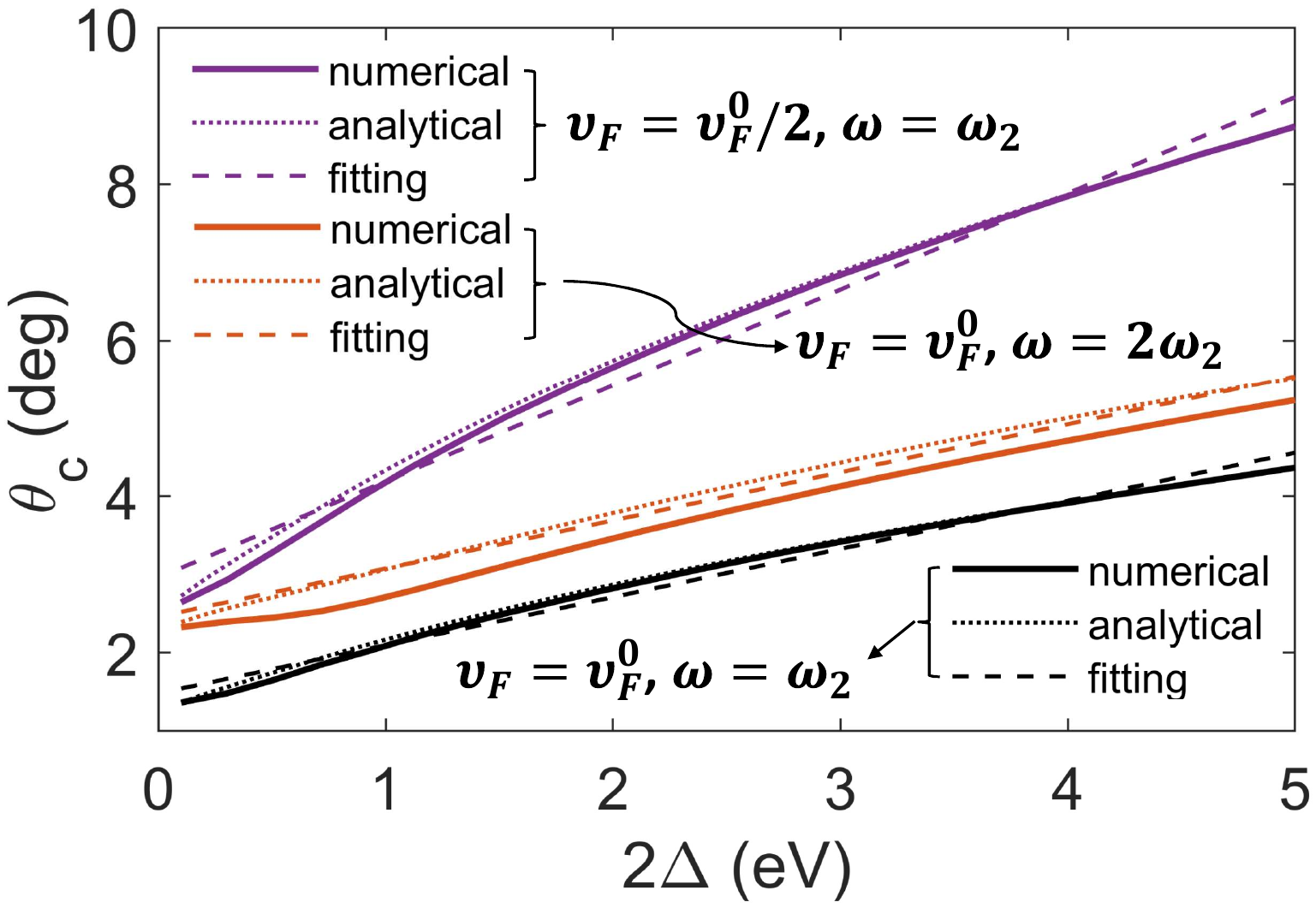}
		\caption{(Color online) Critical twist angle values $\theta_c$ to achieve (a) $W=0.02$~eV and (b) $W=0.03$~eV as a function of the 
		band gap $2\Delta$ obtained from the numerical results, the analytical bandwidths obtained from the first harmonic approximation discussed in Sec.~\ref{analyticalsec}, 
		and line fits of the numerical data in Eq.~(\ref{eq1}).	Here $\upsilon_F^0=0.84\times 10^6$m/s is the LDA Fermi velocity in a single layer graphene. 
		} 
			\label{figure3}
	\end{center}
\end{figure}
While increase in bandwidth $W$ is expected for large twist angles,
a non monotonic behavior can be present depending on the details of the Hamiltonian like in our example of SiC/SiC low energy valence band.
In all the materials considered showing monotonic increase of $W$ with twist angle 
we observe that the critical twist angle $\theta_c$ required to achieve a given bandwidth $W_c$
increases for increasing $\Delta$ and $\omega_i$ and decreasing $\upsilon_{\rm F}$. 
Hence, we can normally expect that the bandwidth will remain smaller than $W_c$ for all the angles below the critical $\theta_c$.
A plausible alternative definition of  $\theta_c$ not used here
would be the ratio $R_c = U/W_c$ between the Coulomb repulsion and bandwidth. 
The relationship between $\theta$ and the system parameters is captured for the twisted gapped Dirac bilayer model with 
a single tunneling parameter $\omega_i = \omega$ through
\begin{eqnarray}
\theta = \frac{C_D \left| \Delta \right| + C_{\omega} \left| \omega \right| + C_W W }{ \left| t_0 \right|}    \label{eq1}
\end{eqnarray}
where we use the dimensionless constants of $C_D = 1.6$, $C_{\omega} = 26$, $C_{W} = 45$, 
and $t_0$ is the nearest neighbor hopping amplitude in a honeycomb lattice that can be related with the Fermi velocity
of the Dirac Hamiltonian through $\upsilon_{\rm F} = \left| t_0 \right|  \sqrt{3}a / 2\hbar$.
The Eq.~(\ref{eq1}) shows the interdependence of $W$ with the system parameters $\theta$, $\Delta$, $\omega$ and can be used to determine the critical angle $\theta_c$ associated to a specific critical bandwidth $W_c$ that we aim for. 
The Fig.~\ref{figure3} compares numerically calculated $\theta_c$ with approximate analytical forms discussed in Sec.~\ref{analyticalsec} and 
confirms the interdependence of the parameters in Eq.~(\ref{eq1}) relating the twist angle $\theta$, 
interlayer coupling strength $\omega_i$, the Fermi velocity $\upsilon_{F}$ of each layer, 
and the staggered potential $\Delta$ between the sublattices within each layer that gives rise to the intrinsic gaps.
\begin{figure*}[htb!]
	\begin{center}
		\includegraphics[width=15cm]{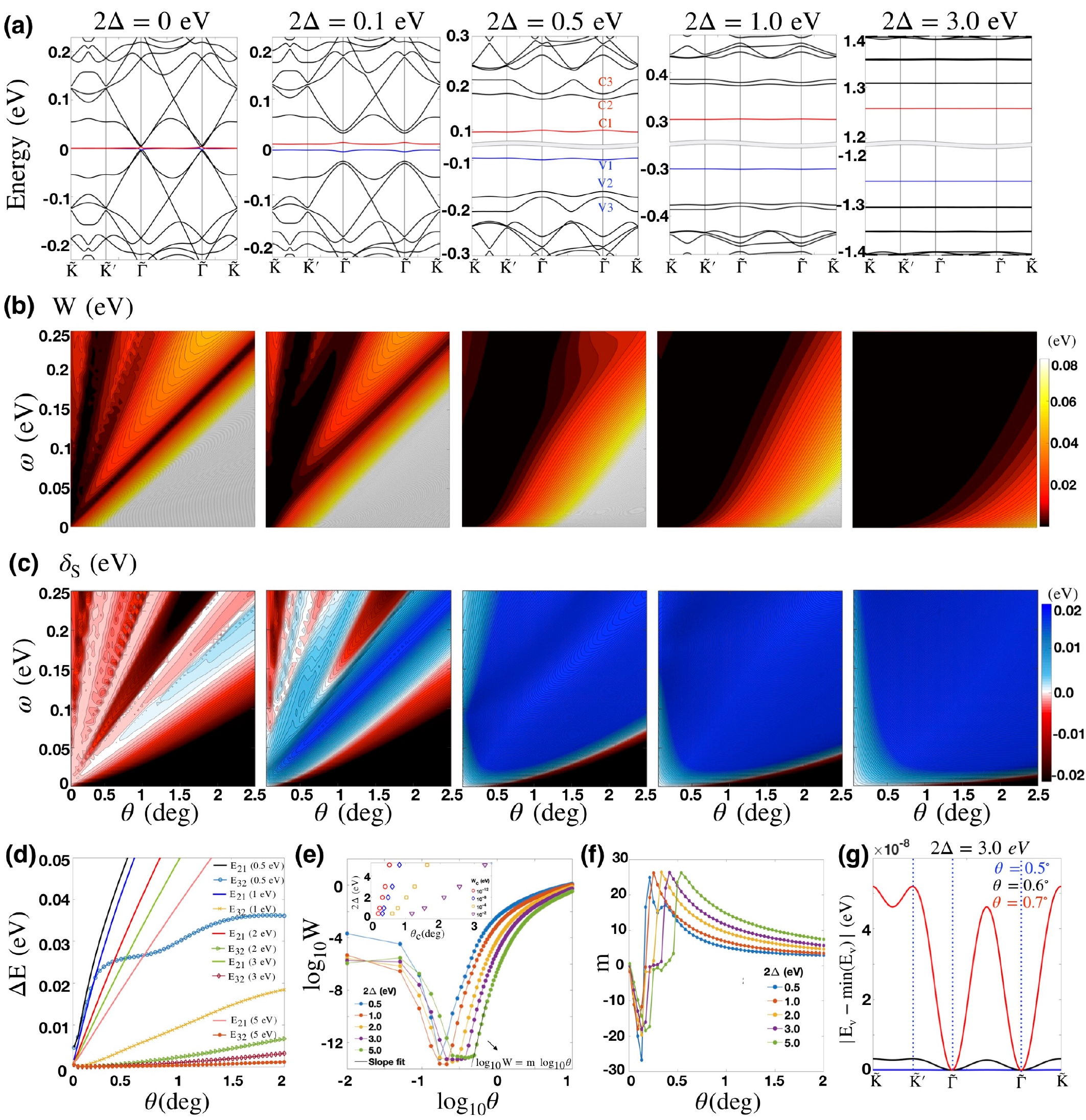}
		\caption{(Color online) 
		(a) Evolution of the electronic structure of $\rm \tilde{G}/\tilde{G}$ at 
		$\theta=1^{\circ}$ twist angle as a function of band gap 2$\Delta=0, 0.1, 0.5, 1, 3$~eV
		where we can observe an enhancement of the secondary isolation gap $\delta_s$ together 
		with the increase in the primary gap $\delta_p$. 
		The band structures have been obtained using equal interlayer coupling parameters 
		$\omega_i = \omega = 0.098$~eV.
		(b) Colormap of the bandwidth $W$ phase diagram in the parameter space of $\omega$ and $\theta$
		where we observe an expansion of the narrow bandwidth region (dark color) when $\Delta$ is increased. 
		(c) Colormap of the isolation gap $\delta_s$ in $\omega$ and $\theta$ phase space where we expect to 
		find isolated flat bands that are	more susceptible to Coulomb interactions. 
		(d) Evolution of the energy spacing between the first three low lying electron-hole symmetric 
		conduction or valence bands evaluated at $\tilde{K}$.
		While the lowest energy band remains approximately constant for different twist angles 
		we find a roughly linear growth in the energy level position with twist angle for the higher energy bands. 
		(e) Bandwidth versus twist angle in a log-log scale shows a bandwidth evolution of the form
		$W({\theta}) \simeq b \theta^{m}$ where the $b(\theta)$, $m(\theta)$ parameters indicates the 
		local evolution of $W$ as a function of $\theta$ that can be fitted with lines in a log-log scale
		for sufficiently large twist angles, whereas the plateaux and non-monotonic behavior for 
		small twist angles are limitations in numerical accuracy. 
		The inset illustrates the almost linear relationship between 2$\Delta$ and critical 
		$\theta_c$ corresponding to fixed values of bandwidth $W_c$. 
		(f) Evolution of the slope $m(\theta)$ for the tangent 
		line fits in panel (e) that decrease progressively for larger twist angles 
		following the hyperbolic dispersion of a single gapped Dirac band.
		(g) Illustration of the rapid change in the bandwidth for nearly flat bands on the order of 
		nano-eV for small variations in the twist angle of $\sim0.1^{\circ}$.
		}
		\label{figure4}
	\end{center}
\end{figure*}

Our calculations show that the MTBG is advantageous for the generation of flat bands in comparison 
to TBG for two important reasons.
First there is no need to aim for a specific magic twist angle to achieve narrow bandwidths when the gaps are sufficiently large,
and second the gap opening allows to achieve flat bands for larger twist angles than in TBG where the structural stability of the moire pattern is greater~\cite{jung2015}.
Flat bands for larger twist angles implies in turn that stronger correlated phases are in principle achievable because the moire Coulomb 
interactions $U \sim  \theta e^2 / 4\pi \varepsilon_r \varepsilon_0 a$ are roughly proportional to twist angle, where $e$ is the electron charge and $a$ is the lattice constant of the individual triangular lattice. 
%
%

This MTBG model is a well defined model that can illustrate the behavior of various gapped Dirac material combinations both for the small and large gap limits.
In the small gap limit a massive graphene layer has experimental realization in graphene aligned with hexagonal boron nitride (G/BN)~\cite{pablo_massive2013,shuyun2016,woods2014,jung2015,suyong2015,ulloa2012,sanjose3,yankowitznature}
that opens a band gap of $\sim 15$~meV or larger depending on sample preparation methods. Likewise
band gaps of the order of few tens of meV are expected in silicene, germanene layers due to intrinsic spin-orbit coupling 
effects~\cite{yuguiyao2011}.
The massive twisted bilayer graphenes can also serve as approximate models for other gapped Dirac materials 
such as SiC, hBN and TMDC~\cite{valleytmd} type triangular lattice 2D materials.

The evolution of MTBG flat bands in massive twisted bilayer graphenes are clearly summarized in Fig.~\ref{figure4} 
where we show the resulting band structure, bandwidth and isolation gaps expected as a function of gap size $2\Delta$, 
twist angle $\theta$, and interlayer coupling $\omega$ for a wide range of parameters that can 
describe several twisted gapped Dirac materials. 
In Fig.~\ref{figure4}(a) we show the band structure from electron-hole symmetrized 
Bistritzer-MacDonald model of twisted bilayer graphene that uses a single interlayer tunneling 
parameter $\omega_i = \omega$ and removes the rotation phases in each graphene layer.
The results are obtained for a fixed twist angle $\theta = 1^{\circ}$
and variable gap sizes between $2\Delta = 0$ till $2 \Delta = 3$~eV where it is clearly shown that finite 
values of intralayer gaps creates a band gap at the primary Dirac point $\delta_p >0$ and also generates
a secondary isolation gap $\delta_s > 0$ whose magnitude increases progressively as $2\Delta$ becomes larger.
It is important that both primary and secondary gaps remain finite in order to 
isolate the low energy bands and allow a stronger effective Coulomb interaction. 
When intralayer band gaps are large enough 
the primary $\delta_p$ and secondary $\delta_s$ gaps open simultaneously
in a larger system parameter space of $\omega$ and $\theta$.
In the limit of small intralayer gaps of a few tens of meV like in aligned G/BN structures this reduced 
bandwidth region concentrate around the magic angle lines in the parameter space and thus we can still 
identify vestiges of the zero gap limit. 
A progressive increase in the gap size alters the phase diagram of the bandwidth in $\omega$ vs $\theta$ space 
merging the discrete traces of magic angle lines into a larger area in the phase diagram expanding
the parameter space where we can find finite secondary gaps $\delta_s$.
This broadening of the parameter space leads to
the disappearance of the discrete magic angles and gives rise to a continuous range of twist angles 
with reduced bandwidth when the band gaps are larger than a few hundreds of meV.
We have typically explored twist angles from $0.05^{\circ}$ to $2.5^{\circ}$ for $\rm \tilde{G}/\tilde{G}$, 
where the tildes indicates the presence of gaps,
and up to $5^{\circ}$ for larger gap materials, 
and interlayer coupling $\omega$ values that span between 0 to 0.25~eV, 
over two times larger than $\omega \sim 0.1$~eV in bilayer graphene. 
For models with unequal $\omega_i$ tunneling values we define $\omega = \omega_2$ and 
keep the same fixed ratios of $\omega_2/\omega_1$ 
and $\omega_2/\omega_3$ when we scale the strenght of interlayer coupling for the different 
materials listed in Table~\ref{table1} 
using $\omega / \omega_i = 1$ for rigid $\rm \tilde{G}/\tilde{G}$,  
$\omega_2/\omega_1 = 1.22$ for relaxed $\rm \tilde{G}/\tilde{G}$,  
$\omega_2$/$\omega_1$ = 0.83, $\omega_2$/$\omega_3$ = 1.89 for BN/BN(0$^{\circ}$), 
$\omega_2$/$\omega_1$ = 0.71, $\omega_2$/$\omega_3$ = 1.89 for BN/BN(60$^{\circ}$), 
and $\omega_2$/$\omega_1$ = 2.5, $\omega_2$/$\omega_3$ = 6.5 for SiC/SiC. 

An additional effect we observe from the gap increase is the progressive flattening of the higher energy 
bands, where we observe bandwidths smaller than a few meVs giving rise to a practically discrete set of 
atomic or quantum dot like energy spectra when individual layer primary gaps assume values of a few eVs
for twist angles around $\sim 1^{\circ}$. 
The twist angle dependence of these quasi-flat higher energy bands follows an approximately linear evolution 
with $\theta$ and is summarized in Fig.~\ref{figure4}(d), 
where we show differences in the energy levels of $E_{21} \sim 40$~meV between the lowest two energy levels,
and smaller than $E_{32} \sim 10$~meV for the energy spacing between the second and third levels when the twist 
angles are changed by $\sim 1^{\circ}$. 
The bandwidth evolution for the higher energy bands in MTBG for different gap sizes are shown in appendix~\ref{hibands}.
Remarkable flattening of the bands down to numerical values of $W \sim 10^{-12}$~eV are represented in Fig.~\ref{figure4}(e)-(g). 
For small twist angles they are widened rapidly with increasing twist angle as $W(\theta) \simeq b \theta^{m}$ 
with powers as large as $m \sim$25 that  gradually reduces with increasing twist angles 
approaching the $m \sim 1$ limiting behavior of hyperbolic bands in the small gap limit. 
This rapid compression in $W$ for sufficiently small $\theta$ implies that the density of states around the average band energy
$E_{\rm av} \sim (E_{\rm max} + E_{\rm min} )/2$ of a given band
\begin{eqnarray}
D(E_{\rm av}) = \int_{\rm mBZ} \frac{d {\bm k}}{(2 \pi)^2} \, \delta( E_{\rm av} - E({\bm k})) \propto \theta^{2} / W \propto \theta^{2-m}
\label{dosevol}
\end{eqnarray}
will roughly be proportional to the moire Brillouin zone area that scales with $\theta^2$
and is inversely proportional to the bandwidth $W$ in the absence of van Hove singularities.
According to this estimate we expect nearly flat bands prone to Coulomb interaction driven ordered 
phases whose bandwidths evolve with twist angle as $D(E_{\rm av}) \propto \theta^{2-m}$ 
even if we do not have divergences associated to saddle points in the Fermi surface.
The rapid enhancement in the joint density of states between the flat bands, when not limited by the broadening due to disorder, 
will more than compensate the decrease of the oscillator strengths in the flat bands to enhance the optical absorption in the system.

%
\begin{figure*}[htb!]
	\begin{center}
		\includegraphics[width=17cm]{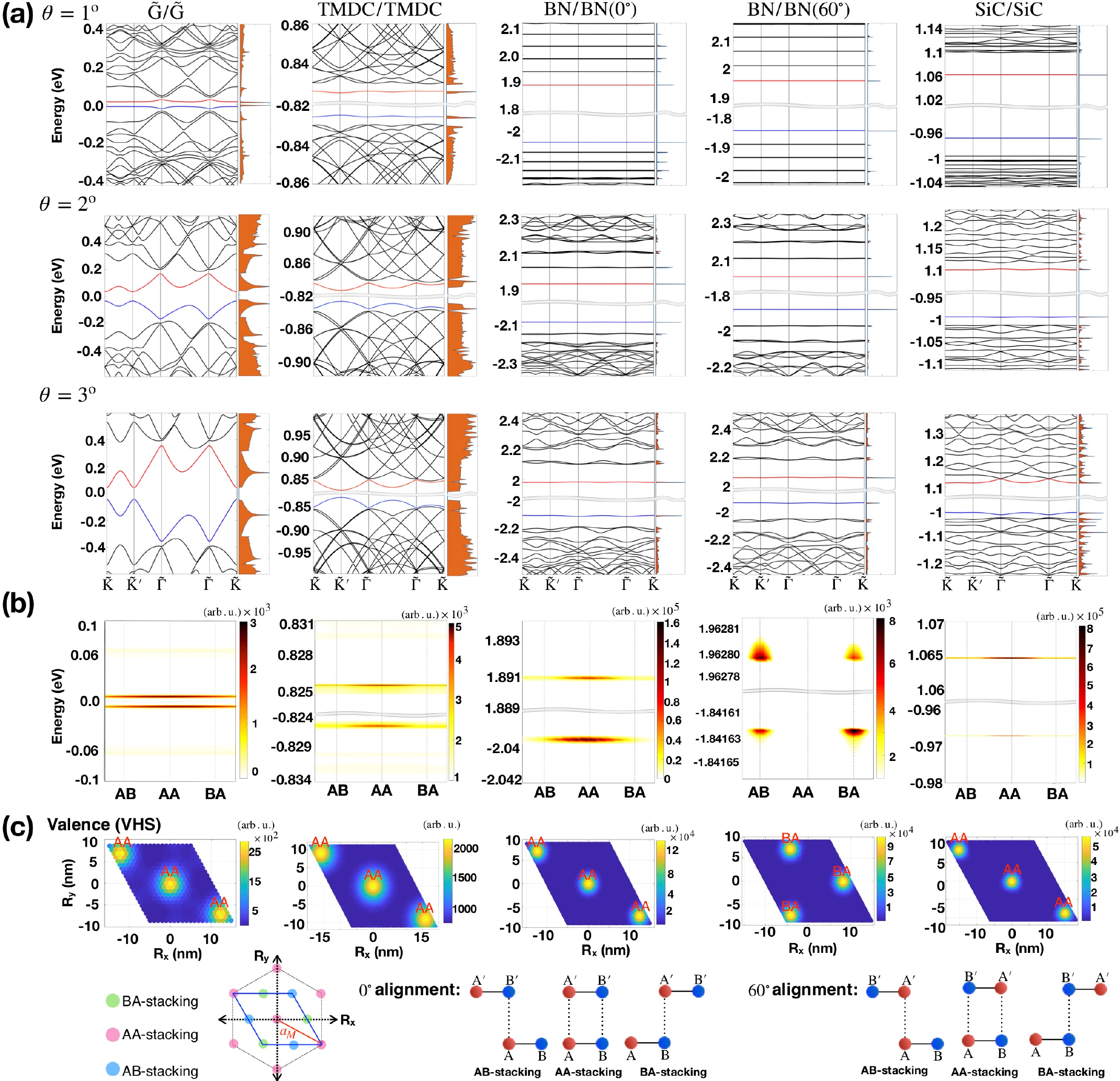} 
		\caption{(Color online) 
		(a) Electronic band structures and the corresponding density of states showing the evolution of flat bands as a function of twist angle 
		in gapped Dirac material bilayer systems. Here, the $\rm \tilde{G}/\tilde{G}$ consist of bilayers whose single layer gap is $2\Delta$ = 0.1~eV. 
		We notice the flattening of higher energy bands as the gaps become larger.
		(b) The local density of states in the bilayer systems at $\theta = 1^\circ$, and 
		(c) In the top panel we show the stacking resolved real space representation of the local density of states at the valence band van Hove singularity (VHS), 
		and in the bottom panel we represent the local stacking configuration of the moire pattern, where the blue lines delimit the LDOS plot area of the top panel,
		and the stacking types in gapped bilayers with $0^\circ$ and $60^\circ$ alignments. 
		}
			\label{figure5}
	\end{center}
\end{figure*}
\begin{figure*}[htb!]
\begin{center}
\includegraphics[width=18cm]{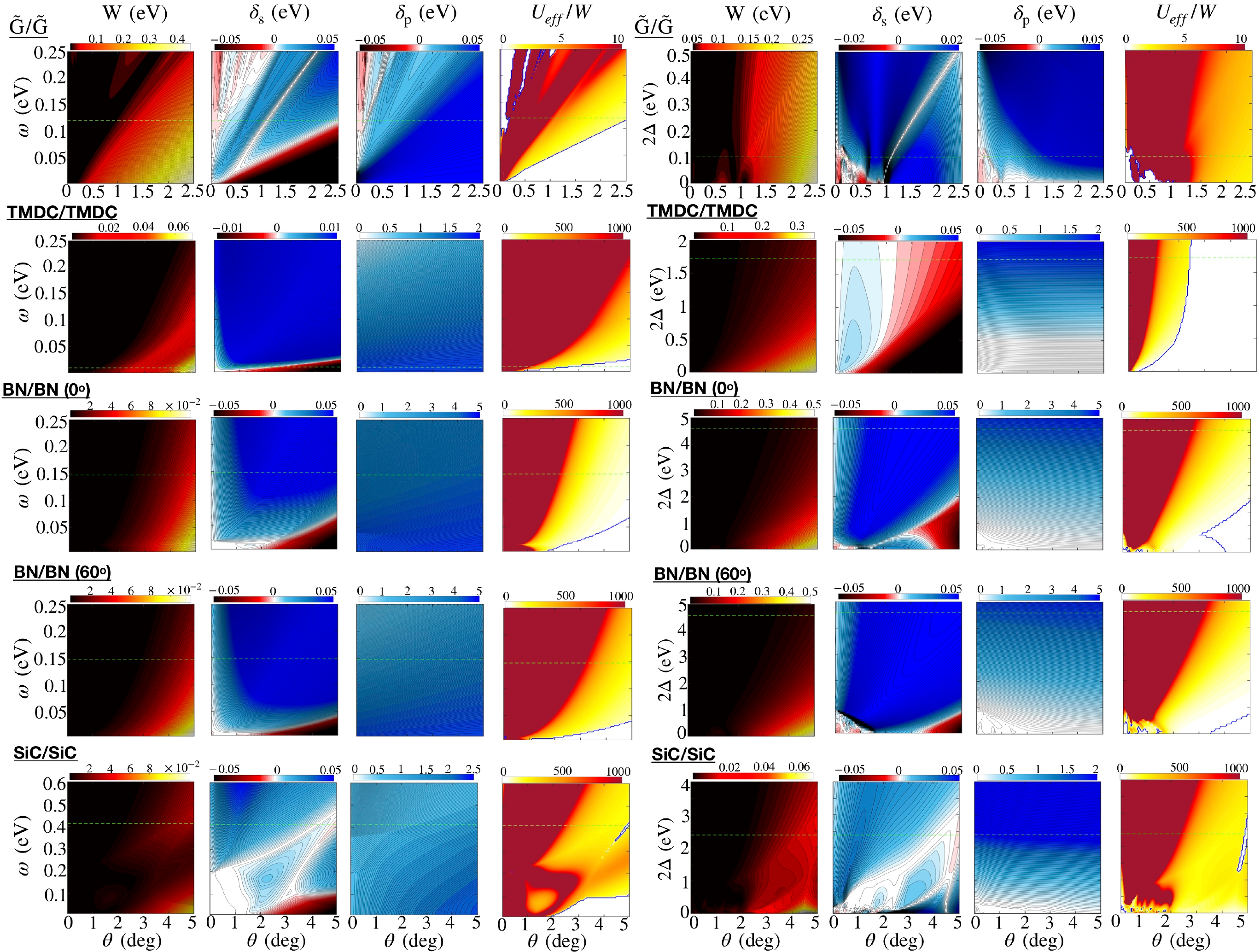}
\caption{
(Color online)  Phase diagram colormaps of the bandwidth $W$, secondary gap $\delta_s$, 
primary gap $\delta_p$, the $U_{\rm eff}/W$ ratio in $\rm \tilde{G}/\tilde{G}$, MoS$_2$/WS$_2$ TMDC, BN/BN and SiC/SiC bilayers. 
The intralayer gaps considered in $\rm \tilde{G}/\tilde{G}$ is of $2\Delta = 0.1$~eV in the first row shows the broadening of the magic angle 
lines that were sharply defined in the limit of zero intralayer gaps 
making the bandwidths less sensitive to twist angle, especially as the intralayer gaps increase. 
Likewise, increase of interlayer tunneling strength generally favors the flattening of the bands. 
Here, $\omega = \omega_2$ is the y-axis, while the remaining $\omega_1$ and $\omega_3$ parameters are proportional to $\omega_2$
with fixed ratios following Table \ref{table1}, as explained in the main text. 
The $\omega_2$ and 2$\Delta$ of each system which are obtained from DFT are indicated with a green horizontal line in each phase diagram. 
The threshold value of the $U_{\rm eff}/W$ = 1 ratio where Coulomb interactions are expected to be significant is indicated with a blue line. 
In systems with large band gaps we find suppressed bandwidths in a large parameter space of twist angle versus interlayer tunneling.
}
\label{figure6}
\end{center}
\end{figure*}

The twist angle dependence of the electronic structure and DOS for a variety of materials and the associated density of states (DOS) and local density of states (LDOS)
are shown in Fig.~\ref{figure5}. 
We notice that when the primary gap $\delta_p$ near charge neutrality and the secondary isolation gap $\delta_s$ are simultaneously present
the low energy flat bands are isolated from the neighboring higher energy bands and they can acquire a well defined integer valley Chern number~\cite{song,senthil,chittari}.
As we commented earlier on, the electronic structure of large gap MTBG like in BN/BN bilayers whose individual layer 
gaps are above several eV shows that higher energy bands are also flattened.
These high energy flatband states should be accessible through carrier gating techniques for 
marginally twisted BN bilayers when the electron densities for each moire band and the interband 
energy spacing between the contiguous flat bands are reduced. 
%

A different perspective for the interdependence of bandwidth and system parameters in $\rm \tilde{G}/\tilde{G}$ 
is seen in the parameter space of $\Delta$ and $\theta$ in Fig.~\ref{figure6}
where we can observe how the traces of the magic angles are erased 
for 2$\Delta \gtrsim$0.2~eV, consistent with the observations in Fig.~\ref{figure2}.
Apart from the obliteration of the magic angles, the increase in the single layer bandgap 
gives rise to a nonzero primary gap $\delta_p$, generally smaller than $2\Delta$, 
that lifts the band degeneracy at charge neutrality,
and opens a secondary isolation gap $\delta_s$ even when all three interlayer tunneling parameters $\omega_i = \omega$ are the same. 
We also observe that the increase of interlayer coupling allows to achieve flat bands at greater twist angles as noted in earlier works~\cite{chittari,stephen2018},
suggesting that we can look for materials with stronger interlayer coupling as a pathway to achieve stronger correlated phases due to reduced moire pattern period, 
as evidenced in our SiC/SiC proposal.

\begin{figure}[t]
	\begin{center}
		\includegraphics[width=6cm]{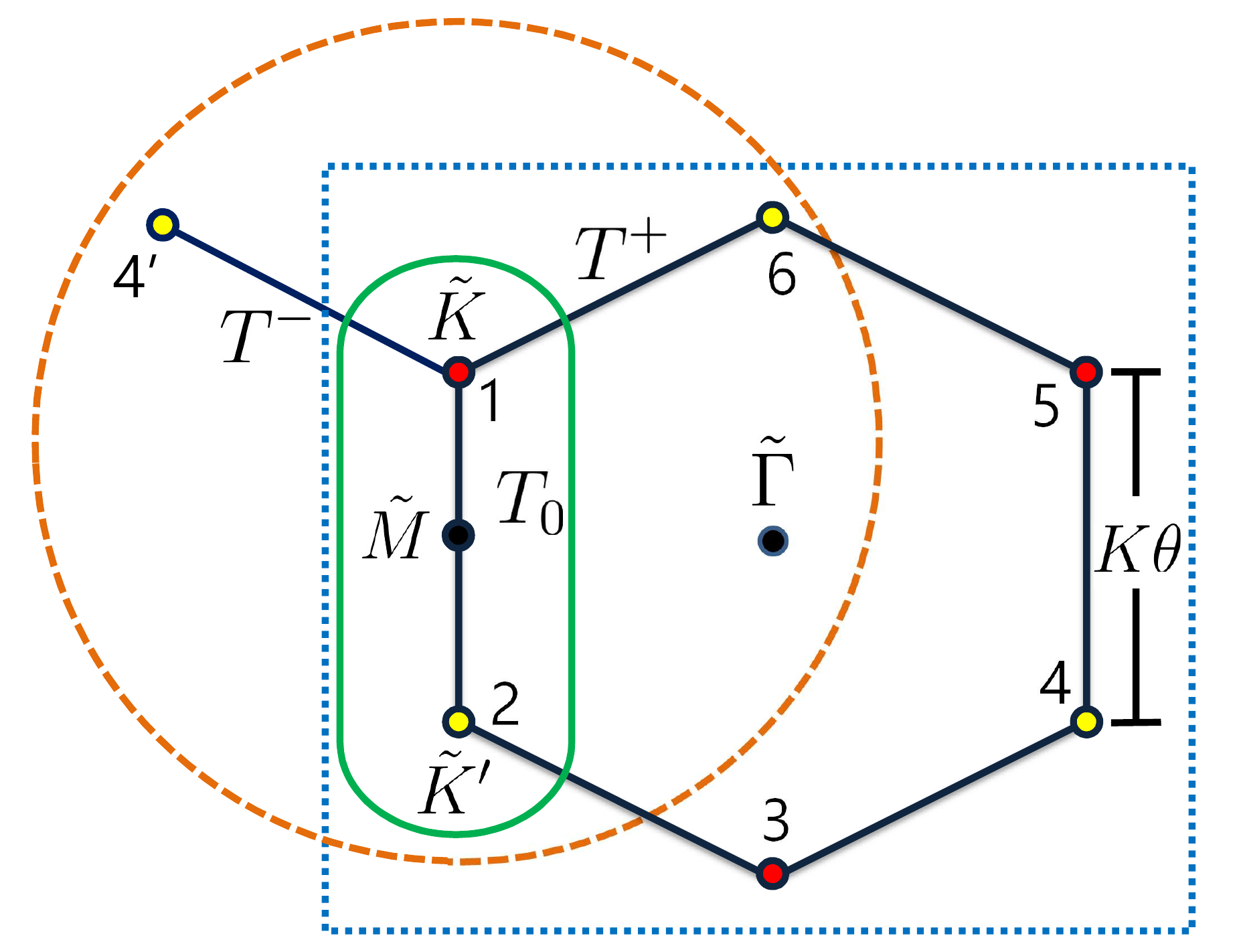}
	\end{center}
	\caption{(Color online) Schematic of high-symmetry points in the mBZ. 
	The numbers $1,\hdots,6$ and $4^\prime$ are the gapped Dirac cone indices, 
	and the red and yellow circles mark the gapped Dirac points from two different layers. 
	$E(\tilde{K})$ is evaluated from a 8-band model consists of four gapped Dirac cones in the dashed brown circle. 
	$E(\tilde{\Gamma})$ is calculated from a 12-band model which includes six gapped 
	Dirac cones inside the dotted blue square. $E(\tilde{M})$ can be obtained from two 
	gapped Dirac cones in the green solid oval. 
	The interlayer tunnelings among the contiguous cones are $T^0$, $T^{\pm}$ 
	as given in in Eq.~(\ref{Eq:Tmatrix}), and we discuss the case $\omega_1=\omega_3\neq \omega_2$.
	}
	\label{figure7}
\end{figure}

Since the primary gap $\delta_p$ near the charge neutrality and the secondary gap $\delta_s$ near the $\Gamma$ point of the mBZ are important factors 
that can influence the onset of interaction-driven phase ordering, we have studied these isolation gaps in the parameter space of ($\theta$, 2$\Delta$) and ($\theta$, $\omega$) for all the massive systems listed in Table.~\ref{table1}. Greater band isolation should reduce screening due to electrons in neighboring bands and therefore strengthen the effective Coulomb interactions.
We have represented in Fig.~\ref{figure6} the $U_{\rm eff}/W$ ratio obtained using the screened Coulomb potential~\cite{tbbg} 
\begin{eqnarray}
U_{\rm eff} = \frac{e^2 }{ 4\pi \varepsilon_r \varepsilon_0 l_M } \exp({- l_M / \lambda_{\rm D}})
\label{ueffeq}
\end{eqnarray} 
where the moire length is $l_M = a/\theta$, and the Debye length
$\lambda_{\rm D} = 2 \varepsilon_0/ e^2 D(\delta_p,\delta_s)$ uses the 2D  density of states
$ D(\delta_p,\delta_s) =  4\, \left( \left| \delta_p \right| u(-\delta_p) + \left| \delta_s \right| u(-\delta_s) \right) / (W^2 A_M) $ 
that assumes a value proportional to the band overlap ratio $\delta_{p/s}/W$ when $\delta_{p/s} < 0$,
where $u(x)$ is the heaviside step function. We use the relative dielectric constant $\varepsilon_r = 4$, 
and there are four valley-spin degenerate electrons per moire unit cell area $A_M = \sqrt{3} \, l_M^2/2$ for each filled moire band.
The phase diagram for $U_{\rm eff}/W$ of gapped $\rm \tilde{G}/\tilde{G}$ systems in the parameter space of $\theta$ and $\omega$ shows that 
for small gaps we can still identify traces of the magic angle lines present in gapless tBG~\cite{chittari2019} 
together with the suppression of the bandwidth and isolation through $\delta_p$ and $\delta_s$ gaps, whereas in large gapped systems we 
can find a larger continuous parameter space of narrow band widths with large $U_{\rm eff}/W$ ratio. 

\section{Analytical expressions of $W$, $\delta_s$, $\delta_p$ and $\delta_M$.}
\label{analyticalsec}
Analytical expressions of the physical quantities relevant for describing the flat bands
including the bandwidth $W$, the primary gap $\delta_p$, the isolation gap $\delta_s$
and avoided gap $\delta_M$ can be obtained from the eigenvalues at the different symmetry 
points in the moire Brillouin zone by solving the truncated moire bands Hamiltonian in the first shell approximation~\cite{jung,bistritzer}. 
The electronic structure of twisted Dirac bilayers results from the coherence between the constituent layers, ranging from the perturbative 
weakly coupled regime for large twist angles to progressively stronger coupling at small twist angles where multiple momenta scattering 
is required in order to capture the electronic structure of the coupled bilayer.
The truncation for the Hamiltonian that we use is therefore just appropriate to describe the systems with large enough twist angles
where the narrow bandwidth starts to become more dispersive. 
%
%
%
%
%
For the analysis in this section we consider the minimal model of the gapped Dirac cones 
connected by the interlayer tunnelings matrices given in Eq.~(\ref{Eq:Tmatrix}) 
and preserve the convenient electron-hole symmetry eliminating the rotation phases 
$e^{\pm i\theta/2}$ phases that accompany the twists in each layer and present our analysis for the 
conduction flat bands which can be equally applied for the valence bands.
We assume for simplicity $\omega_1=\omega_3$ in the tunneling terms and allow for a different $\omega_2$ 
which enhances the isolation gap $\delta_s$.  
We denote by $E_i({\bm k})$ the $i^{th}$ conduction band. 
By using $E_1({\bm k})$ and $E_2({\bm k})$ the band energies of the first and second conduction bands we can define
\begin{eqnarray}
W &=& E_{1}(\tilde{\Gamma}) - E_{1}(\tilde{K})  \label{bandwidth}   \\
\delta_s &\simeq&  E_{2}(\tilde{\Gamma}) - E_{1}(\tilde{\Gamma})  \label{isolationgap}  \\
\delta_p &=&  2 E_{1}(\tilde{K})    \label{primarygap}  \\
\delta_M &=&  E_{2}(\tilde{M}) - E_{1}(\tilde{M})   \label{mpointgap}
\end{eqnarray}
following the properties of the moire bands shown in Fig.~\ref{figure1}(b).
The $E_1(\tilde{K})$ conduction band minimum of the flat band is always at the $\tilde{K}$ point, 
whereas the maximum $E_1(\tilde{\Gamma})$ always happens at the $\Gamma$ point.
Consequently, the bandwidth of the conduction band ($W$) and the primary gap ($\delta_p$) are obtained from $E_1(\tilde{K})$ and $E_{1}(\tilde{\Gamma})$. 
The minimum of the second conduction band is $E_{2}(\tilde{\Gamma})$ for small $\theta$, so the isolation gap is correctly given by Eq.~(\ref{isolationgap}) 
for sufficiently small twist angles when $E_2(\tilde{\Gamma})$ is a minimum. 
For large twist angles, the second conduction band minimum moves away from the $\tilde{\Gamma}$ point starting to deviate from 
Eq.~(\ref{isolationgap}) but this estimate can still be useful for discriminating the isolation of the flat bands.
A comparison of our analytical expressions against numerically obtained $\delta_s$ is offered in the appendix~\ref{analytical}. 
The avoided gap $\delta_M$ or the eigenenergies at $E_{i}(\tilde{M})$ give additional information about the band structure at the zone boundary of the moire Brillouin zone. 

In Fig.~\ref{figure7} we show the schematic of high-symmetry points in the mBZ which are used to obtain the analytical expressions, where the $j=1,\hdots,6$ and $j=4^\prime$  
odd and even indices are used to label the gapped Dirac cones of the bottom and top layers respectively. 
The eignvalues $E_i(\tilde{K})$ are evaluated from an 8-bands model consisting of four gapped Dirac 
cones in the dashed brown circle,
while the $E_i(\tilde{\Gamma})$ are calculated from a 12-bands model which includes six gapped 
Dirac cones inside the dotted blue square. 
The $E_i(\tilde{M})$ energies for the avoided gaps can be obtained from a 4-bands model of two intersecting 
gapped Dirac cones within the green oval in Fig.~\ref{figure7}. 
The interlayer tunneling among the contiguous cones are $T^0$, $T^{\pm}$ as given in Eq.~(\ref{Eq:Tmatrix}), 
and here we limit the discussions to the case of $\omega_1=\omega_3\neq \omega_2$. 
The displacement between the contiguous Dirac points is $K_{\theta} = K \theta$, 
where $K = 4 \pi / 3a$ is the distance between $\Gamma$ to $K$ in the single layer BZ.
The analytical resolution of the eigen energies and eigenvectors for moire systems should also be useful to 
understand the topological properties of the band structures from point group symmetry considerations~\cite{bernevig2012}.

\subsection{ Calculation of $E(\tilde{K})$ from an $8\times 8$ model}
We consider an 8-bands model including only the four gapped Dirac cones inside the dashed circle given in Fig.~\ref{figure7} 
to derive the analytical form of $E(\tilde{K})$. 
Our approach is similar to the scheme used in Ref.~\cite{bistritzer} to calculate the renormalized Dirac-point band velocity in TBG. 
The 8-bands model Hamiltonian connecting one gapped Dirac cone with three surrounding gapped cones reads
\begin{eqnarray}
H_{8\times 8} (\tilde{K}) = \begin{pmatrix} h^1(\tilde{K} )  & T^0 &  T^+ & T^-  \\
T^0 & h^2(\tilde{K})  & 0 & 0 \\
T^+ & 0 & h^6(\tilde{K})  & 0 \\
T^- & 0 & 0 & h^{4'}(\tilde{K})   \end{pmatrix},
\end{eqnarray}
whose diagonal blocks are gapped Dirac Hamiltonians 
\begin{eqnarray}
h^j(\mathbf k) = \begin{pmatrix} \Delta_j & \hbar \upsilon_{\rm F}| {\bm k} - {\bm q}_j|e^{-i\theta_{{\bm k} - {\bm q}_j}} \\
\hbar\upsilon_{\rm F}| {\bm k}-{\bm q}_j |e^{i\theta_{{\bm k} -{\bm q}_j}} & -\Delta_j \end{pmatrix}, 
\end{eqnarray}
where $\bm k$ is the mBZ momentum, and $\theta_{{\bm k}}$ is the momentum orientation with respect to the $k_x$ axis. 
The $j$ is the gapped Dirac cone index, so ${\bm q}_j = \{q_{x,j},\ q_{y,j}\}$ determines the center of the $j$-th Dirac cone.   
The eigenstate consists of four two-component spinors
$\Psi^T=\{\psi_1^T, \ \psi_2^T, \ \psi_3^T, \ \psi_4^T \}$.   
We find that at the $K$ point, the spinors for the first conduction band always follows the relations 
$\psi_2 = \mathcal{U} \psi_4$, $\psi_3 = \mathcal{U}^* \psi_4$, where $\mathcal{U}$ is a diagonal matrix given by 
\begin{eqnarray}
\mathcal{U} = \begin{pmatrix}  1 & 0 \\
0 & e^{\text{i}2\pi/3} \end{pmatrix}. 
\end{eqnarray}\label{Eq:UforEK}
The discussions for the valence bands are closely similar and they are shown in appendix~\ref{analytical}. 
Consequently, the first conduction band energy at the $\tilde{K}$ point is given by 
\begin{equation}
\begin{aligned}
E(\tilde{K}) &=  \frac{\Delta}{3} + \frac{2}{3} \sqrt{4\Delta^2 +3\rho_\theta^2+9\omega_1^2+9\omega_2^2}\cos(\frac{2\pi-\varphi}{3}), 
\end{aligned} \label{Eq:EK}
\end{equation}
where 
\begin{eqnarray}
\rho_\theta &\equiv& \hbar \upsilon_{\rm F} K \theta    \nonumber \\
\varphi &=& \text{acos}\frac{8\Delta^3 + 9\Delta \rho_\theta^2 -54\Delta\omega_1^2 + 27\Delta\omega_2^2}{(4\Delta^2 +3\rho_\theta^2+9\omega_1^2+9\omega_2^2)^{3/2}}.   \nonumber
\end{eqnarray}
For $\omega_1=\omega_3$, the twisted gapped Dirac bilayer model given in Eq.~(\ref{Eq:Htbg}) 
preserves electron-hole symmetry in the absence of the aforementioned rotation phases, 
and the lowest band edges reside at the $\tilde{K}$ point
Hence the resulting analytical expression of the primary gap is
\begin{equation}
\begin{aligned}
\delta_p = 2E(\tilde{K}),
\end{aligned}
\end{equation}
where $E(\tilde{K})$ is given by Eq.~(\ref{Eq:EK}).
\begin{figure}[tb!]
	\begin{center}
		\includegraphics[width=8cm]{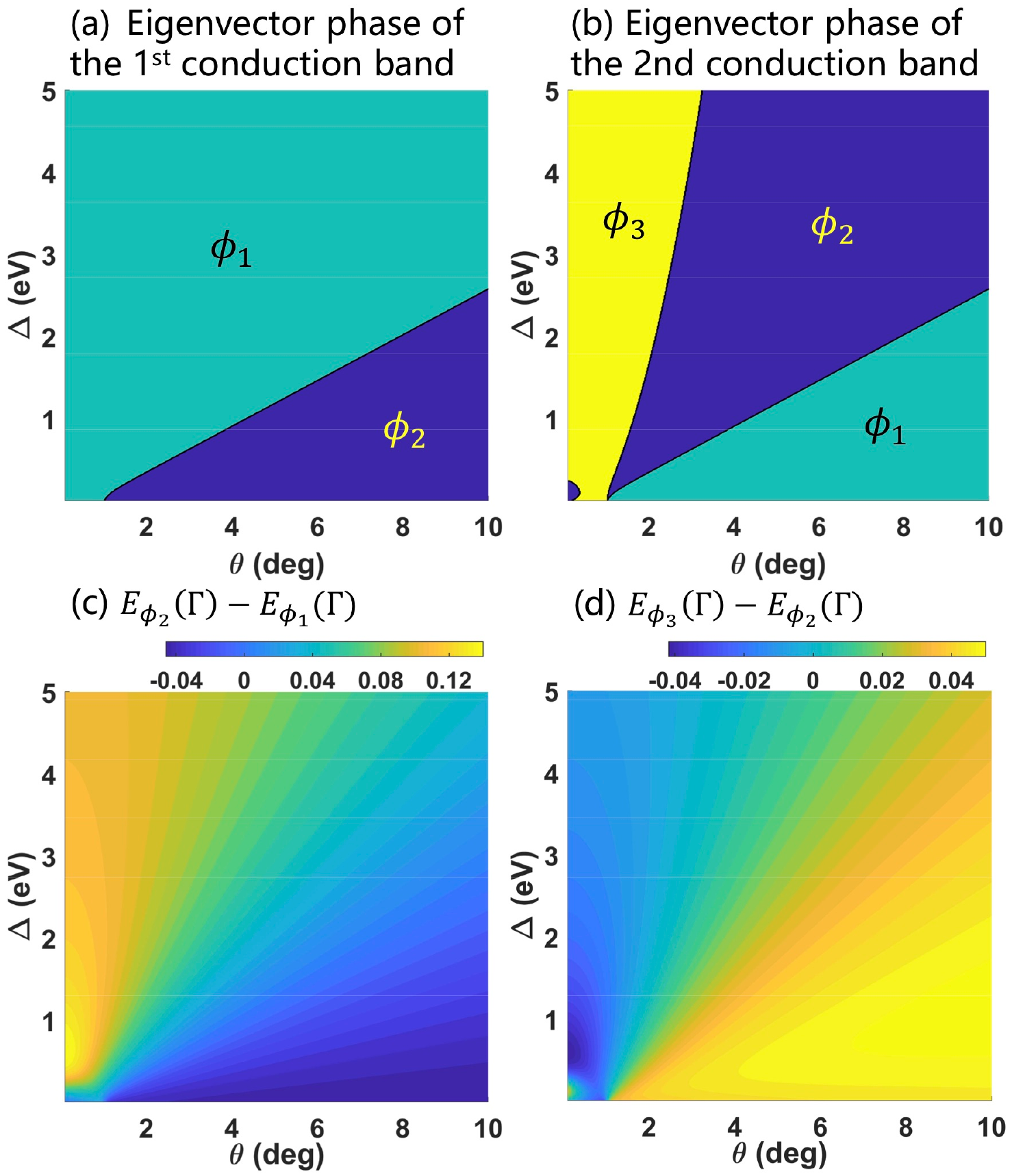}
	\end{center}
	\caption{(Color online). 
	Plots in the top row show the eigenvector phases of the (a) first and (b) second $E_i(\tilde{\Gamma})$ conduction bands, 
	where the $\phi_i$ labels in the plot denote the eigenenergies $E_{\phi_i}(\tilde{\Gamma})$ for
	$\phi_1= \{0, 2\pi/3\}$, $\phi_2= \{2\pi/3, -2\pi/3\}$, and  $\phi_3=\{-2\pi/3, 0\}$. 
	Plots in the second row show the differences between different $E_{\phi_i}(\tilde{\Gamma})$ in the space of $\Delta$ and $\theta$. 
	The parameters used in these calculations are $\omega_1=\omega_2=\omega_3=0.098$~eV and $t_0 = -2.6$~eV. }
	\label{supp_phase}
\end{figure}

Due to the electron-hole symmetry, the first valence band energy at $\tilde{K}$ point 
$E^{\rm v}(\tilde{K}) = - E(\tilde{K})$ is available from Eq.~(\ref{Eq:EK}).
For the valence band states at $\tilde{K}$ we can define the two-component spinors for the $8\times 8$ model
in a similar way to what we have done for the conduction band. 
The eigenstate consists of four two-component spinors $\Psi^T=\{\psi_1^T, \ \psi_2^T, \ \psi_3^T, \ \psi_4^T \}$,
and the first valence band always follows the relations 
$\psi_2 = \mathcal{U}_v \psi_4$, $\psi_3 = \mathcal{U}_v^* \psi_4$, where $\mathcal{U}_v$ is a diagonal matrix given by 
\begin{eqnarray}
\mathcal{U}_v = \begin{pmatrix}  e^{\text{i}2\pi/3} & 0 \\
0 &  1 \end{pmatrix}. 
\end{eqnarray}

\subsection{Analytical expression of $E(\tilde{\Gamma})$} 

The eigenenergy at the $\tilde{\Gamma}$ point can be calculated from the $12\times 12$ Hamiltonian consisting of 
six gapped Dirac cones inside the blue dotted square in Fig.~\ref{figure7}. 
The interlayer tunneling between contiguous gapped Dirac cones are given by the $T_0$, $T^\pm$ matrices, 
and the Hamiltonian reads 
\begin{equation}
\begin{aligned}
&H_{12\times 12} (\tilde{\Gamma}) = \\
&\begin{pmatrix} 
h^1( \tilde{\Gamma}) & T^0 & 0 & 0 & 0 & T^+ \\
T^0 & h^2( \tilde{\Gamma}) & T^- & 0 & 0 & 0 \\
0 & T^- & h^3( \tilde{\Gamma}) & T^+ & 0 & 0 \\ 
0 & 0 & T^+ & h^4( \tilde{\Gamma}) & T^0 & 0 \\
0 & 0 & 0 & T^0 & h^5(\tilde{\Gamma}) & T^- \\
T^+  & 0 &  0 & 0 & T^- & h^6( \tilde{\Gamma}) \\\end{pmatrix}. 
\label{Eq:H12}
\end{aligned}
\end{equation}
The eigenstate consists of six two-component spinors
$\Psi^T=\{\psi_1^T, \ \psi_2^T, \ \psi_3^T, \ \psi_4^T, \ \psi_5^T,\ \psi_6^T \}$,   
and we find that the spinors always follow the relations 
$\psi_3 = \mathcal{V} \psi_1$, $\psi_5 = \mathcal{V}^* \psi_1$, $\psi_4 = \mathcal{V} \psi_2$, $\psi_6 = \mathcal{V}^* \psi_2$, and
$\mathcal{V}$ is a diagonal matrix
\begin{eqnarray}
\mathcal{V} = \begin{pmatrix}  e^{i\eta_1} & 0 \\
0 & e^{i\eta_2} \end{pmatrix}
\label{Eq:eta}
\end{eqnarray}
where $\{\eta_1, \eta_2\}$ only have three different combinations, which are
$\phi_1= \{0, 2\pi/3\}$, $\phi_2= \{2\pi/3, -2\pi/3\}$, and  $\phi_3=\{-2\pi/3, 0\}$. 
The eigenvalues for the valence bands at the $\tilde{\Gamma}$ point have electron-hole symmetry 
and are related to the conduction band energies through $E_{\phi_1}^{\rm v}(\tilde{\Gamma})=-E_{\phi_3}(\tilde{\Gamma})$, 
$E_{\phi_2}^{\rm v}(\tilde{\Gamma})=-E_{\phi_2}(\tilde{\Gamma})$ and 
$E_{\phi_3}^{\rm v}(\tilde{\Gamma})=-E_{\phi_1}(\tilde{\Gamma})$. 
Likewise the valence band eigenvectors can be defined with the same spinor structure for the conduction bands
but using the $\phi_i$ phases that follow from this electron-hole symmetry of the eigenenergies. 
For each $\phi_i$ ($i=$1,2,3), the eigenenergy problem is then changed to solving a quartic equation 
\begin{eqnarray}
x^4 + \alpha x^2 + \beta x + \gamma =0.
\label{Eq:quarticmain}
\end{eqnarray}
For the three $\phi_i$ sets, the coefficients in Eq.~(\ref{Eq:quarticmain}) become different, and by solving the equation for each $\phi_i$ one can get 4 roots, among which only one root is useful in our calculation. Hence for three $\phi_i$s we obtain three different energy eigenvalues and denote them as $E_{\phi_i}(\tilde{\Gamma})$ 
whose exact expressions are given in appendix~\ref{analytical} and here we use the following approximations
\begin{widetext}
	\begin{equation}
	\begin{aligned}
	E_{\phi_1 / \phi_3}(\tilde{\Gamma}) & \approx \sqrt{\Delta^2 +\rho_\theta^2+2.5\omega_1^2 + \omega_2^2 -\sqrt{10\Delta^2\omega_1^2 + 9\rho_\theta^2\omega_1^2 +9/4\omega_1^4 + 4 \rho_\theta^2\omega_2^2 + \omega_1^2\omega_2^2}  }\\ 
	& \mp \frac{3\Delta\omega_1^2}{2\sqrt{10\Delta^2\omega_1^2 + 9\rho_\theta^2\omega_1^2 +9/4\omega_1^4 + 4 \rho_\theta^2\omega_2^2 + \omega_1^2\omega_2^2}},\\ 
	E_{\phi_2}(\tilde{\Gamma})&=\sqrt{\Delta^2 + \rho_\theta^2 + \omega_1^2 + 4\omega_2^2 -2\sqrt{\Delta^2\omega_1^2+4\rho_\theta^2\omega_2^2+4\omega_1^2\omega_2^2}}  
	\end{aligned}
	\end{equation}
\end{widetext}
We represent in Fig.~\ref{supp_phase} the $\phi_i$ parameters that determine the 
eigenstates of the first and second conduction flat bands.
In this figure the first conduction band energy is given by either 
$E_{\phi_1}(\tilde{\Gamma})$ or $E_{\phi_2}(\tilde{\Gamma})$, 
while the second conduction band energy is given by either $E_{\phi_1}(\tilde{\Gamma})$, $E_{\phi_2}(\tilde{\Gamma})$ 
or $E_{\phi_3}(\tilde{\Gamma})$ depending on the Hamiltonian parameters $\Delta$ and $\theta$.
Consequently, the expressions of the bandwidth and the isolation gap are 
\begin{equation}
\begin{aligned}
W &= \min\left(E_{\phi_1}(\tilde{\Gamma}), E_{\phi_2}(\tilde{\Gamma})\right) - E(\tilde{K}), \\
\delta_s & = |\text{min}(E_{\phi_2}(\tilde{\Gamma}), E_{\phi_3}(\tilde{\Gamma})) - E_{\phi_1}(\tilde{\Gamma})|.    
\end{aligned}
\end{equation}


\subsection{Analytical models for $E(\tilde{M})$}
\begin{figure}[tb!]
	\begin{center}
		\includegraphics[width=8.5cm]{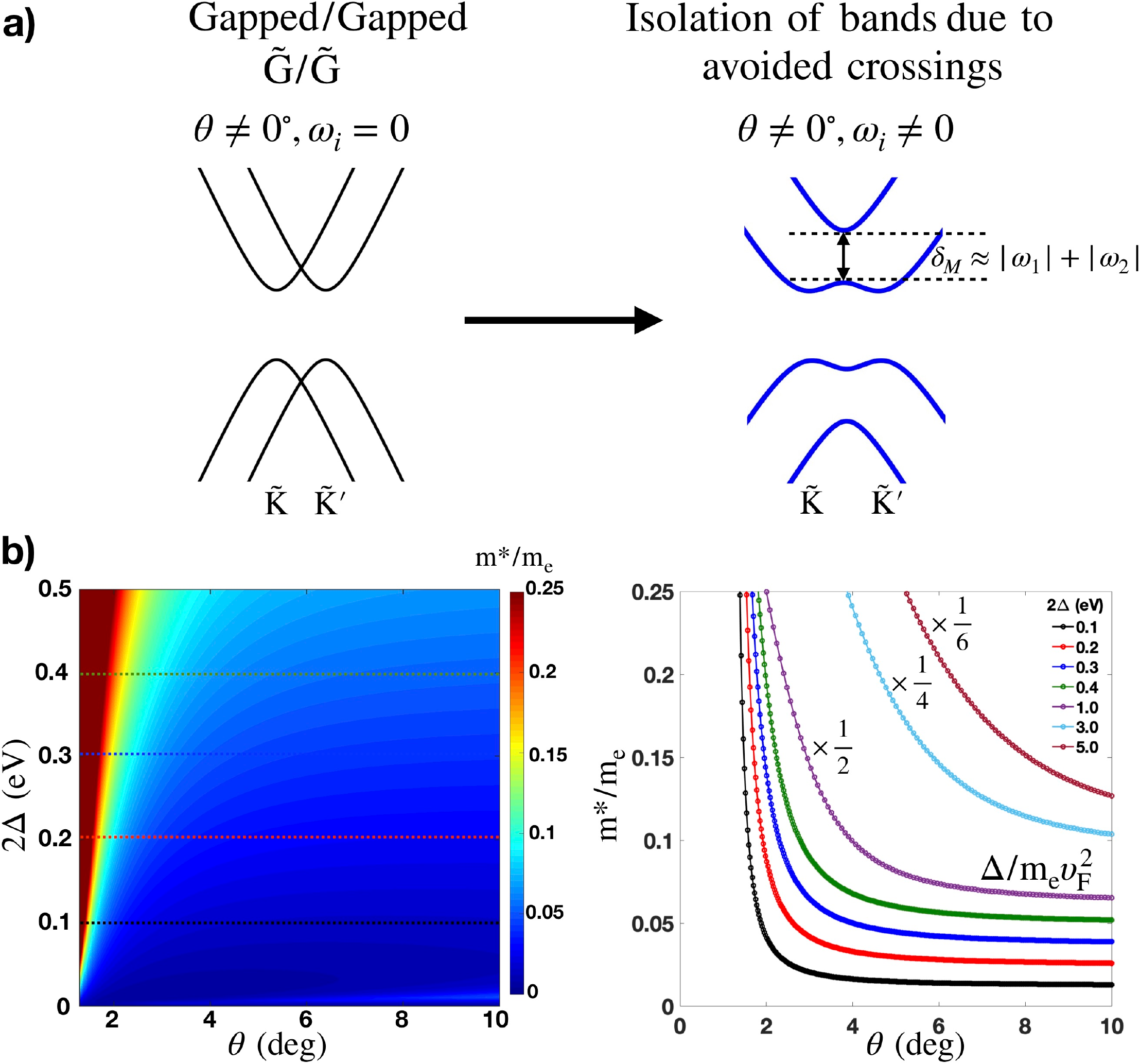} 
		\caption{(Color online)  Effective mass $m^{*}/m_{\rm e}$ with respect to the rest mass of the electron 
		 as a function of $2\Delta$ and $\theta$  for the massive twisted bilayer graphene. In the left panel, the effective mass is obtained from the numerical derivation of Hamiltonian in Eq.~(\ref{Eq:Htbg}) at the band edge. In the (right) panel, we show line cuts that illustrate the divergence of the effective mass near a critical twist angle that depends on $\Delta$,
		that for large enough twist angles when the band edges locate at $\tilde{K}$ and $\tilde{K}'$
		approaches $m^* = \Delta/\upsilon_{\rm F}^2$.
				}
		\label{effectivemass}
	\end{center}
\end{figure}

The moire bands in reciprocal space are repeated periodically by the moire reciprocal lattice vectors ${\rm G}_m$.
The coherence between the two overlapping bands of the rotated layers can be approximated in the simplest perturbative limit by 
a four by four Hamiltonian neglecting all but the smallest momenta coupling equivalent to two interacting gapped Dirac cones 
where the bands would otherwise intersect between $\tilde{K}$ and $\tilde{K}'$ as illustrated in Fig.~\ref{effectivemass}.

\begin{equation}
\begin{aligned}
H_{4\times 4} (\tilde{M}) = 
\begin{pmatrix} 
h^1( \tilde{M}) & T^0 \\ 
T^0 & h^2( \tilde{M}) 
\\\end{pmatrix}. 
\label{Eq:H4}
\end{aligned}
\end{equation}

The energy eigenvalue at the $\tilde{M}$-point can be obtained by diagonalizing Eq.~(\ref{Eq:H4}). 
For the simplest case when $\omega_1=\omega_2=\omega_3$, 
the two conduction band energies at the $\tilde{M}$-point are given by 
\begin{equation}
\begin{aligned}
E(\tilde{M}) = \sqrt{\Delta^2 + (\rho_\theta/2)^2 + \omega_1^2} \pm  \left| \omega_1 \right|,
\end{aligned}
\end{equation}
where $\rho_{\theta} = \hbar \upsilon_{\rm F} K \theta$
from which we can conclude that the energy difference for the avoided gap is
\begin{equation}
\begin{aligned}
\delta_M = 2 \left| \omega_1 \right|.
\label{deltam1}
\end{aligned}
\end{equation}
If $\omega_1=\omega_3\neq\omega_2$, the $\tilde{M}$-point conduction band energy eigenvalues are 
\begin{equation}
\begin{aligned}
&E(\tilde{M}) =\\
& \sqrt{\Delta^2 + (\rho_\theta/2)^2 + \omega_1^2 + \omega_2^2 \pm 2\sqrt{\Delta^2\omega_1^2+ 
	(\rho_\theta/2)^2\omega_2^2 +\omega_1^2\omega_2^2}}. 
\end{aligned}
\end{equation}
In the limit of large $\Delta$, one can approximate $E(\tilde{M})$ to obtain
\begin{equation}
\begin{aligned}
\delta_M \approx 2\frac{\sqrt{\Delta^2\omega_1^2+ (\rho_\theta/2)^2\omega_2^2 + \omega_1^2\omega_2^2}}{\sqrt{\Delta^2 + (\rho_\theta/2)^2 + \omega_1^2 + \omega_2^2 }},
\end{aligned}
\end{equation}
which further simplifies to 
\begin{equation}
\begin{aligned}
\delta_M \approx \left| \omega_1 \right|  +  \left| \omega_2 \right|
\label{deltam2}
\end{aligned}
\end{equation}
if we assume $\omega_2=\omega+\delta\omega$, $\omega_1 = \omega-\delta\omega$ and $\delta \omega \ll \omega_1, \omega_2$.

The above Hamiltonian and associated solutions can be further simplified to a $2 \times 2$ model
in the limit of large band gaps thanks to the almost perfect sublattice polarization of the conduction and valence bands. 
In this limit, we can build a two by two Hamiltonian that couples two conduction or two valence  
mutually coupled through a single tunneling parameter $\omega$,
and  in the presence of an interlayer potential difference $V_g$ we have
\begin{eqnarray}
 \hat{H}_{2 \times 2}( k ) = \begin{pmatrix}   \varepsilon_{c(v)}( k + \frac{K_{\theta}}{2})  + \frac{V_g}{2} & \omega  \\ \omega^{*}  &  
 \varepsilon_{c(v)}( k - \frac{K_{\theta}}{2}) - \frac{V_g}{2}  \end{pmatrix}   
\end{eqnarray}
that couples through $\omega$ the two conduction or valence band edges 
$ \varepsilon(k) = \varepsilon_{c}(k) = - \varepsilon_{v} (k) = 
(  \Delta^2 + (\hbar \upsilon_{\rm F} (k ) )^2 )^{1/2}     $   
arising from the top and bottom layers whose bands represented along the line that connects the two shifted moire zone $\tilde{K}$-points are shown in Fig.~\ref{effectivemass} for $V_g = 0$. 
By denoting $\varepsilon^{\pm}(k) = \varepsilon( k \pm K_{\theta}/2 )$ for the band edges
we obtain the following eigenvalues
\begin{widetext}
\begin{eqnarray}  
E_{c(v)} (k) &=& \left( \frac{\varepsilon^{-}_{c(v)} (k ) + \varepsilon^{+}_{c(v)} (k )}{2} \right) \label{eq2bands} 
 \pm  \sqrt{ \left| \omega \right|^2 + \left(  \frac{\varepsilon^{-}_{c(v)} (k ) - \varepsilon^{+}_{c(v)} (k )-V_g}{2} \right)^2}
\label{Ekanalyt}
\end{eqnarray}
\end{widetext}
Keeping in mind that at the  $\tilde{M}$ point  $\varepsilon^{+}(0)=\varepsilon^{-}(0) = \varepsilon(\pm K_{\theta}/2 )$ the avoided gap reduces to 
\begin{eqnarray}
\delta_M &=& 2 \left| \omega \right|
 \end{eqnarray}
in agreement with Eq.~(\ref{deltam1}) and Eq.~(\ref{deltam2}).
\begin{figure*}[htb!]
	\begin{center}
		\includegraphics[width=18cm]{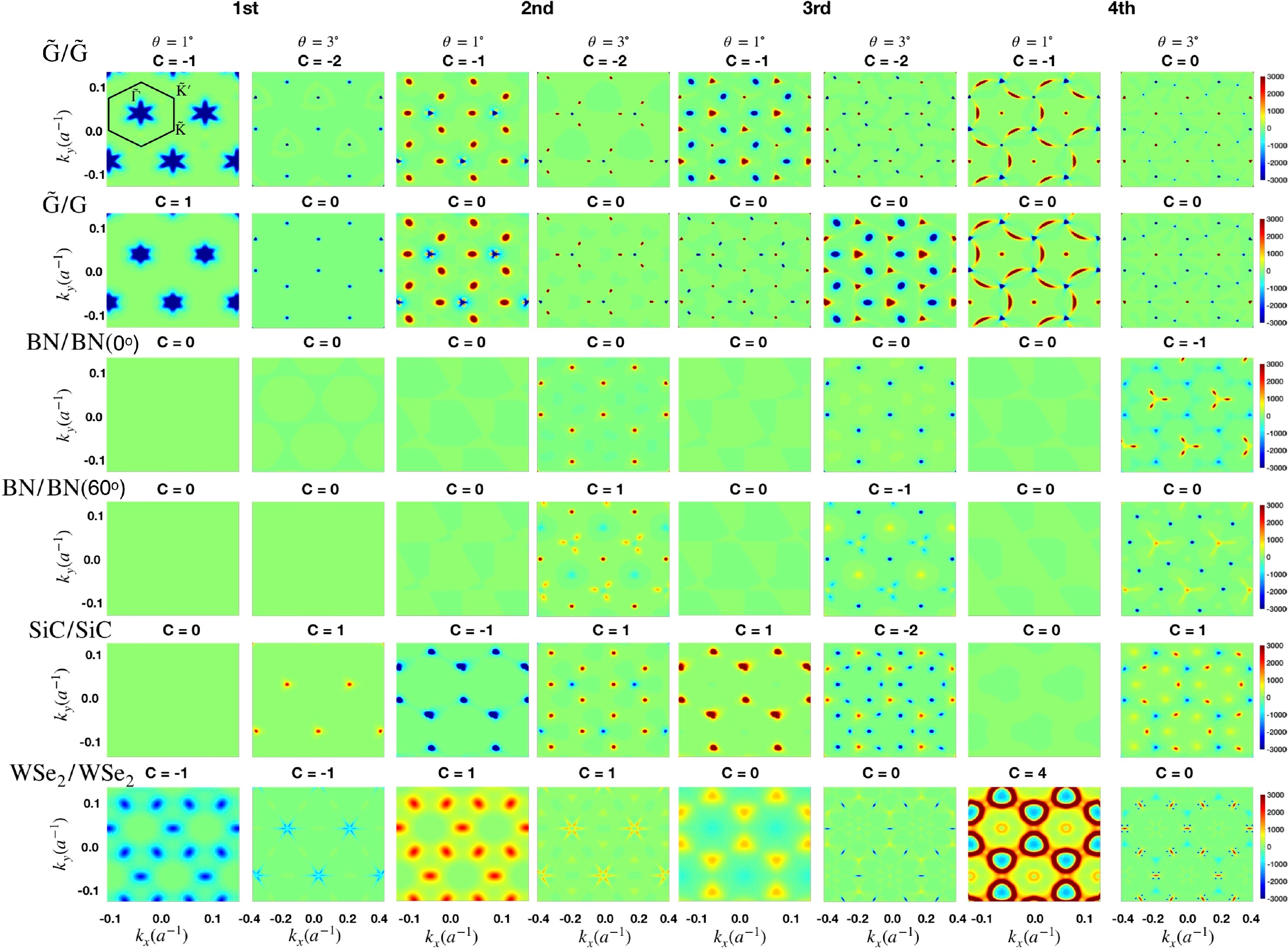}
		\caption{(Color online)  Berry curvatures for the first four conduction bands of different twisted gapped Dirac Hamiltonian bilayers modeled
		through the parameters in Table~\ref{table1}.  The peaks of the Berry curvature are often located at the 
		symmetry points of the moire Brillouin zone where the avoided gaps are produced and their values depend sensitively on the Hamiltonian details.  
		For increasing twist angles the two twisted gapped Dirac Hamiltonians tend to effectively decouple for low energies, 
		and therefore the Berry curvature tend to concentrate more sharply near the symmetry points of the mBZ where the gaps decrease.
		}
		\label{berrycurv1}
	\end{center}
\end{figure*}
The effective mass $m^*$ of the electron given by 
\begin{eqnarray}
m^* = \hbar^2 / \frac{\partial^2E( { k} )}{\partial { k}^2}
\label{mstar}
 \end{eqnarray}
has an analytical approximation near $(\tilde{K},~\tilde{K}^{\prime} )$ given by,
\begin{eqnarray}
m^* = \frac{\varepsilon(k) ^3}{\Delta^2\upsilon_F^2}
\label{largeangle}
\end{eqnarray}
that reduces to the $K$ point value $m^* = \Delta / \upsilon_F^2 $ of a single gapped Dirac cone
in the limit of large twist angles~\cite{xiaodi}.
The numerical $m^*$ from the low energy bands are represented in Fig.~\ref{effectivemass} as a 
function of gap and twist angle.

\begin{figure*}[htb!]
	\begin{center}
		\includegraphics[width=16.5cm]{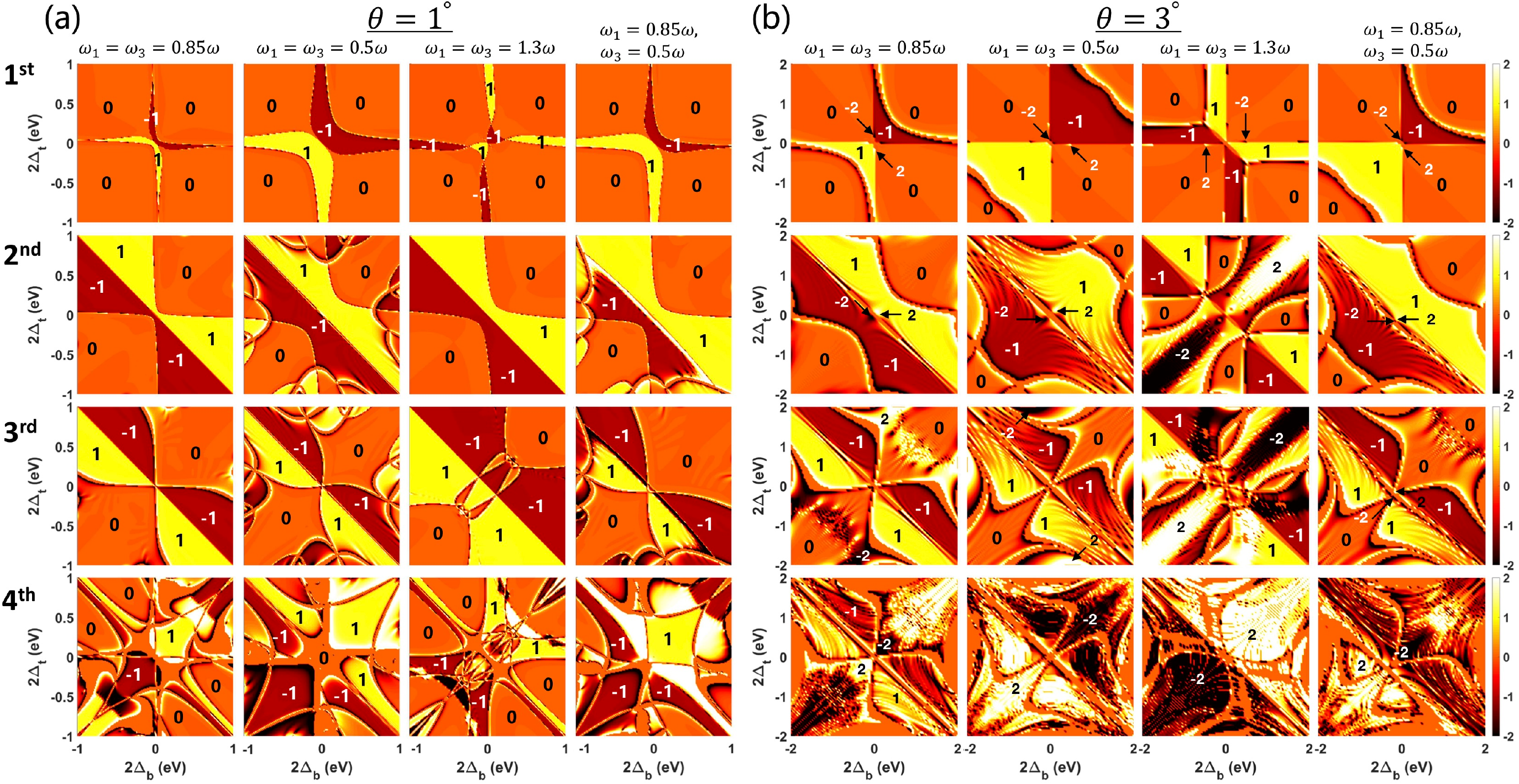}
	\end{center}
	\caption{(Color online)
	Valley Chern number phase diagram of massive twisted bilayer graphene as a function of the mass terms $\Delta_{t}$ and $\Delta_{b}$ for the top and bottom layers for the first four conduction bands. 
	Different models for interlayer coupling are represented in each column.
	We notice that the phase space of non-trivial Chern bands increases with the differences in magnitude of the tunneling $\omega_i$ between different sublattices. 
	For large $\Delta$ values, the increase of twist angle helps to expand the parameter space region that turns the trivial low energy bands into Chern bands.
	The rapid variations in the Chern numbers in the high energy bands indicate that presence of multiple band crossings that prevents the formation of well developed gaps between the neighboring bands. 
		}	
	\label{figure11}
\end{figure*}
\begin{figure*}[tb!]
	\begin{center}
		\includegraphics[width=16.5cm]{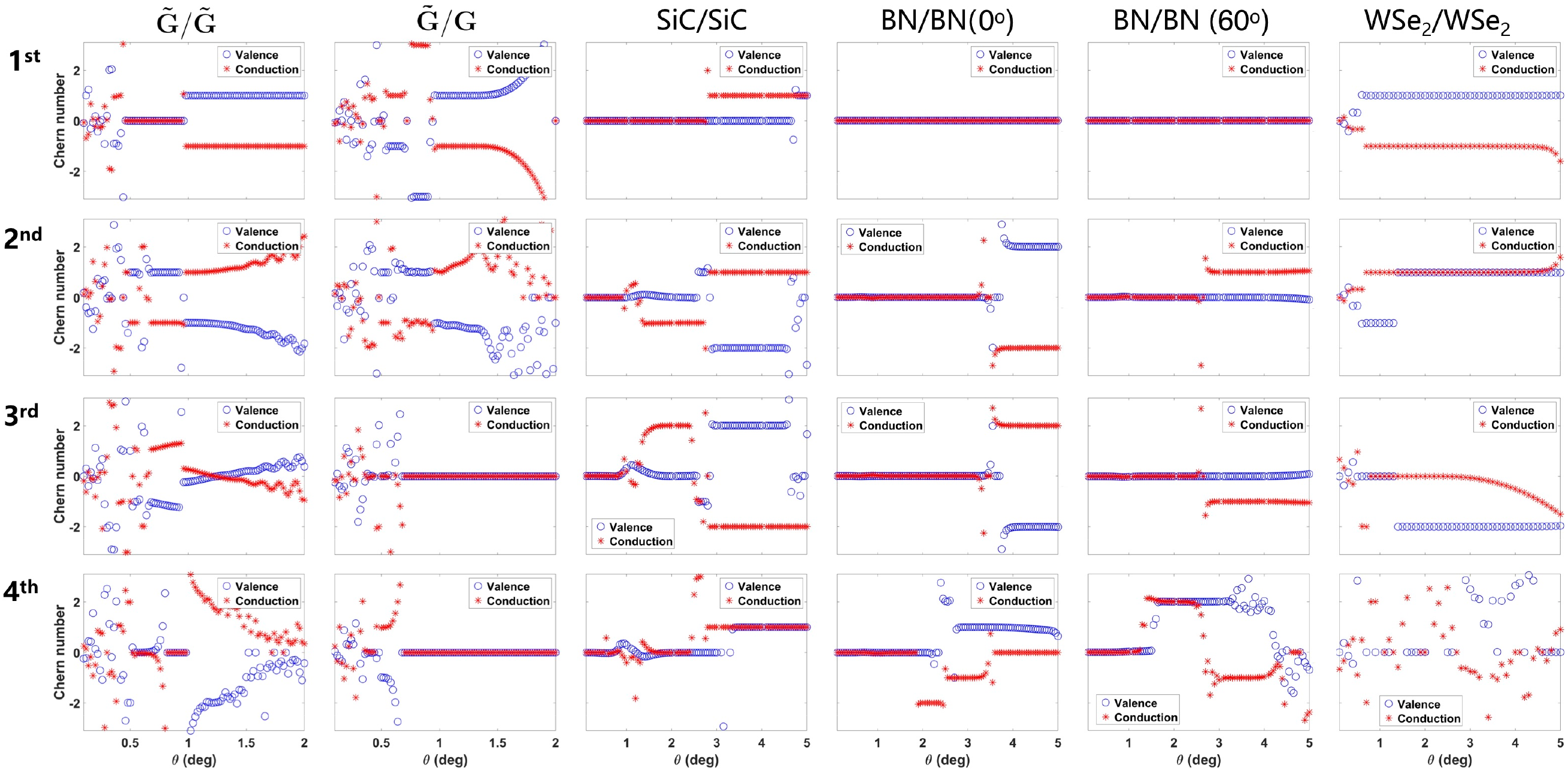}
	\end{center}
	\caption{(Color online)
	Chern number results for the first four conduction and valence bands calculated with respect to $\theta$ for a variety of twisted massive Dirac cone systems 
	modeled through the parameters in Table.~\ref{table1}.
	The $\rm \tilde{G}$ represents an intrinsically gapped graphene with $2\Delta = 0.01$~eV of comparable magnitude of the gap in G/BN, while $\rm G$ represents a gapless graphene.
	We find relatively wide twist angle regions where the Chern numbers are well quantized especially for the lower energy bands.  
	When the bands are not well isolated the valley Chern numbers are not quantized to an integer value 
	and can show rapid variations when the system parameter changes. 
	}	
	\label{chern_angle}
\end{figure*}

\section{Topological flat bands and valley Chern numbers}
\label{chernnumbers}
Isolated bands in the moire Brillouin zone can develop integer valley Chern numbers 
different from zero depending on the details of the Hamiltonian~\cite{song,chittari2019,senthil2019}.
A finite charge Hall conductivity will develop when the valley degeneracy is lifted by valley polarization in an 
otherwise time reversal invariant system~\cite{senthil,chittari}.
This type of selective valley and spin filling mechanisms 
have been proposed to underlie the observed ferromagnetism 
and anomalous Hall effects in twisted bilayer graphene aligned with a boron nitride substrate~\cite{mott3,andreayoung2019},
and the {\rm ordered phases in twisted double bilayer graphene~\cite{tbbg1,tbbg2,tbbg3,tbbg4}.} 
On a closely related topic for our work, opposite valley Chern numbers of $C = \pm1$ have been proposed in twisted bilayer graphene with a small gap of $\sim$15~meV in one of the 
layers aligned with the hexagonal boron nitride substrate~\cite{zaletel,senthil} while Coulomb 
interactions could trigger $C = \pm 2$ low energy bands even in the absence of substrate effects~\cite{macdonald2019}.
Analogous valley Chern bands due to intralayer moire patterns have also been proposed in twisted semiconducting TMDC~\cite{fengcheng2019}. 
Recent observations of quantized anomalous Hall conductance at zero magnetic field in TBG~\cite{andreayoung2019} 
or non quantized orbital moments~\cite{mott3},
and at small magnetic fields of $\sim 0.4$~T in ABC trilayer graphene on hexagonal boron nitride
in the presence of electric field induced gaps of $\sim 20$~meV~\cite{fengwang2019} suggest 
optimistic prospects of finding exotic zero magnetic field quantum Hall states when the device qualities are 
sufficiently improved.

In the following we show the valley Chern number phase diagram expected in MTBG for different sets of system parameters
$\Delta$, $\upsilon_F$, $\theta$, $\omega_i$ ($i=$1,2,3) and $V_g$ for the first four conduction bands near charge neutrality
expanding beyond the parameter region studied in earlier related work~\cite{senthil,zaletel,chittari}, see appendix~\ref{massdep} for a similar phase diagram 
and Berry curvature plots of the valence bands. 
We calculate numerically the valley Chern numbers through the standard formula~\cite{berry_rmp}
\begin{equation}
\begin{aligned}
C_{\nu} = \int_{\text{mBZ}} \text{d}{\bm k} \ \Omega_n({\bm k})/2\pi
\end{aligned}
\end{equation}
where $\nu = \pm 1$ is the valley index, using the Berry curvature of the $n^{th}$ band  
\begin{equation}
\begin{aligned}
\Omega_n({\bm k})=-2 \sum_{n\neq n'}\text{Im}\left[\frac{\langle u_n| \frac{\partial H}{\partial k_x}|u_{n'} \rangle  
\langle u_{n'} |\frac{\partial H}{\partial k_y}|u_n\rangle}{\left(E_n-E_{n'} \right)^2}\right]
\end{aligned}
\end{equation}
where $E_n$ and $u_n$ are the eigenenergies and eigenvectors.  
The maxima of Berry curvatures as illustrated in Fig.~\ref{berrycurv1} 
often concentrate around the different symmetry points of the mBZ at $\tilde{K}$, $\tilde{K}'$ or $\tilde{\Gamma}$, 
and near the moire Brillouin zone boundaries where the avoided gaps are formed. 
Some general conclusions we can anticipate from our calculations of the Chern number phase diagram
of twisted gapped Dirac bilayers are that (i) the conduction and valence bands remain topologically 
trivial if $\omega_1=\omega_2=\omega_3$ and their relative magnitudes sensitively contribute in determining the phase 
space of topological bands, (ii) as the band gap of the system becomes larger we need to increase the 
twist angle $\theta$ to turn the lowest band into a Chern band, (iii) 
the intralayer moire patterns contribute in the determination of the Chern number phase diagram, and
(iv) small values of interlayer bias comparable in magnitude with the interlayer tunneling can lead 
to higher energy band crossings and changes in Chern number, as was seen in models of TMDC
consisting of intralayer moire potentials~\cite{fengcheng}.

In Fig.~\ref{figure11} we represent the Chern number phase diagrams for the first four conduction bands of two twisted gapped Dirac materials as a function of the gap magnitudes.
The phase diagrams for the valence bands are similar to that of conduction bands with opposite Chern numbers except for electron-hole symmetry breaking 
and they are presented in appendix~\ref{massdep}.
We have shown the phase diagrams corresponding to different twist angles of $\theta = 1^{\circ}$ and  $\theta = 3^{\circ}$ 
in Fig.~\ref{figure11}'s (a) and (b) panels respectively where we find that shortening of the moire period by increasing $\theta$ 
can turn the low energy bands into Chern bands in large gap systems. 
This behavior can be explained if we consider that an increase in twist angles enlarges the mBZ area and 
prevents the Berry curvature weights at the low energy bands to be pushed to higher energies. 
Sensitive changes of Chern number as a function of gap magnitude found for large twist angles
and high energy bands indicate the complex avoided crossing structure at higher energies. 
We have listed in each column the Chern bands phase diagram 
for same $\upsilon_{\rm F}$ parameter as in graphene and use different interlayer 
tunneling terms $\omega_1$, $\omega_3$ and variable gap values $\Delta_b$ ($\Delta_t$) of the bottom (top) layer. 
In the first and second columns we show the results for the interlayer tunneling parameters of 
$\omega_1=\omega_3=0.85\omega_2$ and $\omega_1=\omega_3=0.5\omega_2$. 
As noted earlier, if we choose the values of $\omega_i$ ($i=$1,2,3) to be identical, both conduction and valence bands 
are trivial in the $\Delta_b$, $\Delta_t$ phase space, but if the diagonal tunneling terms $\omega_1=\omega_3$ become 
distinct from the off-diagonal $\omega_2$ 
the integer valley Chern number emerges in two belt regions
that become wider as the difference between $\omega_1$ and $\omega_2$ increases.    
Hence, the low energy Chern bands are possible only when there are at least two different $\omega_i$ values
coupling different sublattices in  Eq.~(\ref{Eq:Tmatrix}), and a sufficiently small finite mass term in at least one of the layers. 
The nontrivial Chern band parameter region is generally larger in $\Delta_t, \, \Delta_b$ space
when the differences in the tunneling parameters
are also large, as we can verify comparing $\omega_1 = \omega_3 = 0.85 \omega_2$ and 
$\omega_1 = \omega_3 = 0.5 \omega_2$ columns,
and we have an intermediate situation for $\omega_1 = 0.85 \omega_2, \, \omega_3 = 0.5 \omega_2$ where we additionally include
a difference between the diagonal hopping parameters $\omega_1, \, \omega_2$. 
The former case is closely related with the experimental situation of TBG where one layer is aligned with BN.
When $\Delta_b = \Delta_t$ the maximum allowed gap is $\sim$50~meV before the level becomes trivial, whereas
if $\Delta_t = 0$ the maximum $\Delta_b$ allowed to preserve a topological band is about $\sim$140~meV.
A qualitatively different phase diagram is found when $\omega_1=\omega_3 = 1.3 \omega_2$
when the diagonal tunneling elements become larger than the off-diagonal interlayer terms, 
while we still require small enough mass terms to preserve a Chern band. 
However, even when the mass terms are as large as a few eVs it is possible
to find Chern bands for large enough twist angles and higher energy bands.
The valley Chern bands appear even for small twist angles $\theta \lesssim 1^{\circ}$ in WSe$_2$/WSe$_2$ system 
modeled mainly through intralayer moire patterns.
%
%
%
%
%
%
We have represented in Fig.~\ref{chern_angle} the twist angle $\theta$ dependence of the Chern 
numbers for several massive twisted bilayer graphene systems that we modeled from the parameters in Table~\ref{table1}. 
For small band gaps we find nonzero valley Chern numbers for the lowest energy bands in a wide range of twist angles $\theta \gtrsim 1^{\circ}$ in the limit of small band gaps.
When the intralayer band gaps are larger we notice a tendency for higher energy bands to acquire finite 
Chern numbers for sufficiently large twist angles $\theta \gtrsim 3^{\circ}$ for interlayer tunneling dominated 
moire pattern systems like SiC/SiC and BN/BN bilayers.%
\begin{figure}[tbh!]
	\begin{center}
	\includegraphics[width=8cm]{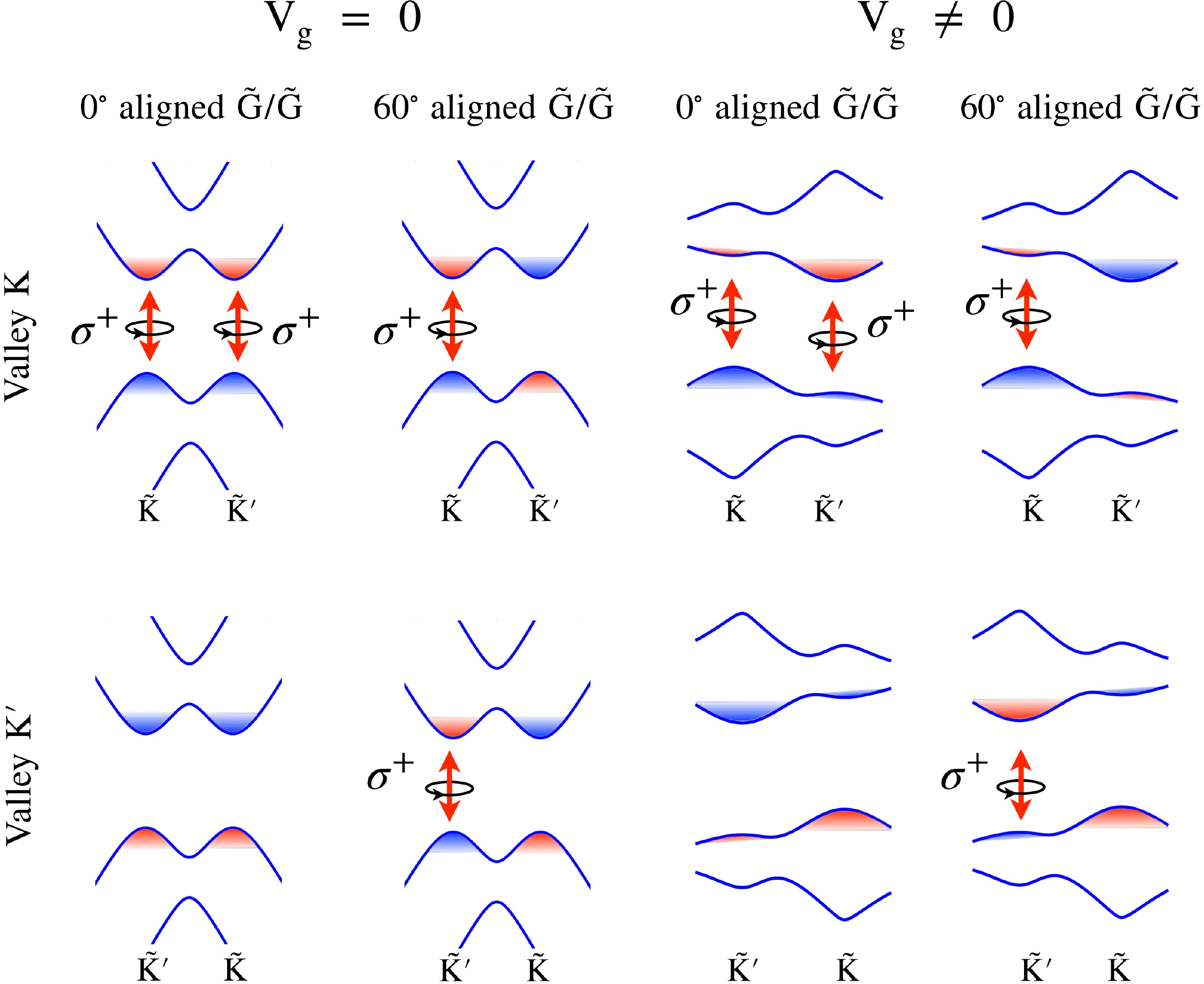}
	\end{center}
	\caption{(Color online). 
	Schematic illustration of twisted gapped Dirac bilayers for zero and finite interlayer potential difference near  $0^{\circ}$
	and $60^{\circ}$ rotation alignments giving rise to same and opposite mass signs respectively at the minivalleys $\tilde{K}$ and $\tilde{K}'$.	We can expect macrovalley $K$ and $K'$ contrasting circular dichroism near $0^{\circ}$ alignment and suppressed circular dichroism near $60^{\circ}$.
	For the latter it is possible to introduce a finite circular dichroism by applying interlayer bias and 
	Fermi level change that leads to layer and minivalley polarization (within a macrovalley).
	}	
	\label{interband}
\end{figure}

\section{Valley contrasting optical transitions}
\label{selectionrules}
Numerous optical experiments for semiconducting transition metal dichalcogenides have verified optical dichroism in broken inversion 
symmetry single layer materials~\cite{xiaodi_exciton, yaowang2007,tcao}
associated with the chirality of the layers in twisted multilayer systems~\cite{jiwoongpark2016,breydichroism,gomezsantos2018}. 
In the simplest picture, circular dichroism is expected in gapped Dirac materials due to valley contrasting 
orbital moments associated with the Berry curvatures at the band edges for circularly polarized dipole optical interband transitions involving 
angular momentum changes of $\Delta l = \pm 1$~\cite{yaowang2007},
and leading to selection rules of the form $m = w \pm 1$  for the promotion of angular momentum $m$ excitons 
in $w$-chiral gapped Dirac systems~\cite{xiaodi_exciton,tcao}.
Here we investigate how in a gapped twisted bilayer graphene system the circular dichroism for the interband optical transitions 
are modified going from the decoupled layers limit consisting of two independent 
gapped Dirac Hamiltonian layers to two coupled gapped Dirac Hamiltonians leading to strongly hybridizing flatbands. 
In our models this crossover in behavior can be achieved either by changing the rotation angle between the bilayers or by modifying the interlayer coupling strength.
\begin{figure}[tb!]
	\begin{center}
		\includegraphics[width=8.6cm]{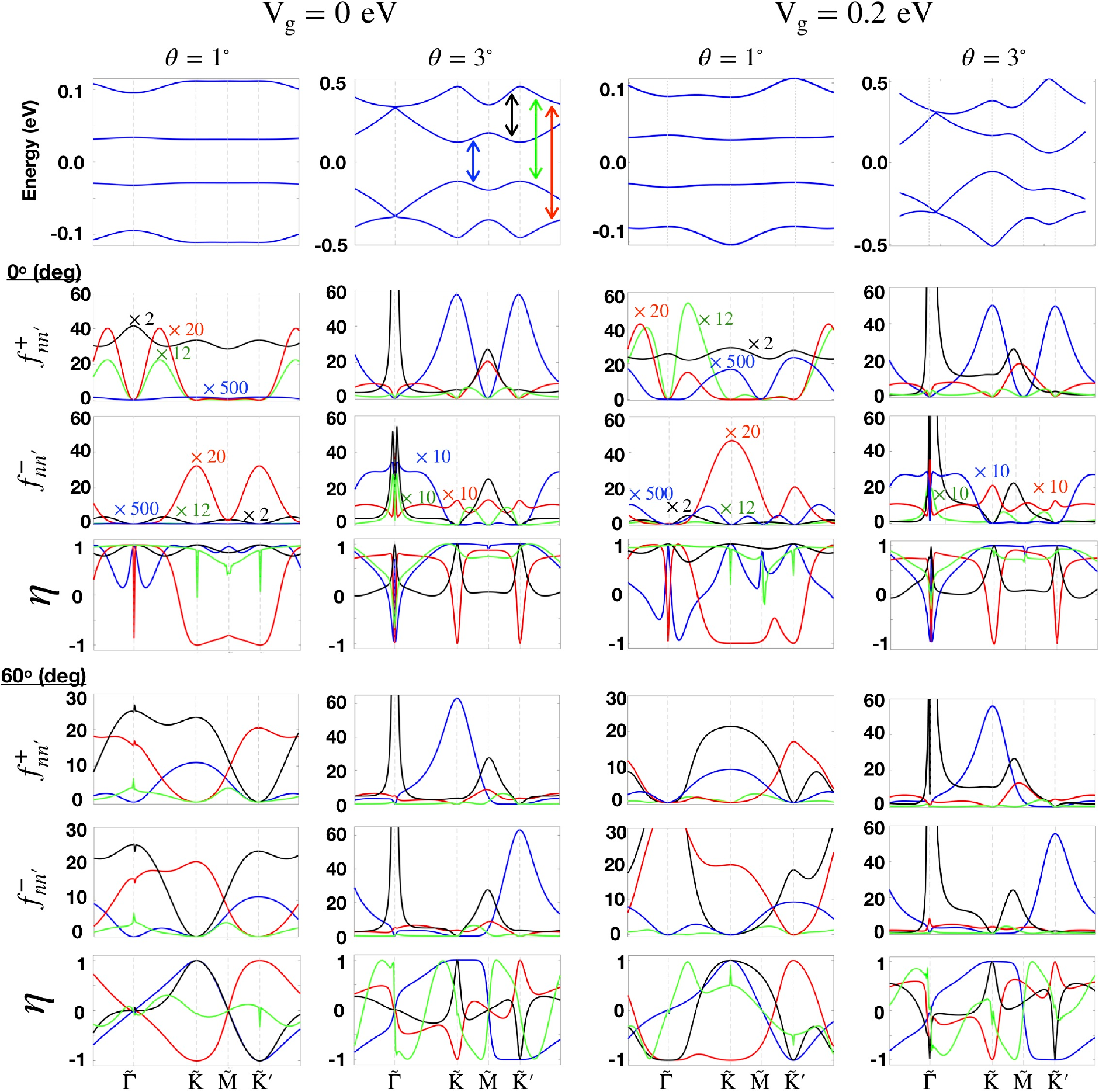}
		\caption{(Color online) 
		Interband transition oscillator strengths $f^{\pm}_{nn'} ({\bm k})$ and dichroism $\eta_{nn'} ({\bm k})$ for circularly polarized light in twisted gapped Dirac material bilayers for various twist angles showing clear peaks around the gapped Dirac cone band edges at $\tilde{K}$ and $\tilde{K}'$. These peaks become broader 	and lower as the twist angle is progressively reduced and the quasiparticle velocities decrease. 		
		The comparison between the twisted bilayers near $0^{\circ}$ and $60^{\circ}$ alignment shows oscillator strength peaks with the same and opposite signs near each minivalley. We observe same sign and opposite sign circular dichroism $\eta ({\bm k})$
		near the minivalleys indicative of the switch from macrovalley to minivalley contrasting physics depending on alignment. 
		The transitions between selected bands are distinguished with blue (1v and 1c), black (1v and 2c), red (1c and 2c) and green (2v and 2c) solid lines.  
		We used the interlayer coupling $\omega_i = 0.098$ ~eV and $\left| t_{0} \right| = $2.6~eV, with a gap of 2$\Delta$ = 0.3 eV.
	}
	\label{interband2}
	\end{center}
\end{figure}
In a single gapped Dirac Hamiltonian model the interband optical oscillator strength for each 
$k$-point for circularly polarized light is given by~\cite{cardona} 
\begin{eqnarray}
f^{\pm}_{ n n'} ( {\bm k} )  
= \frac{2 \left| P^{\pm}_{ n n'}  ({\bm k}) 
\right|^2}{ m_e \hbar \omega_{ n n'}({\bm k})   }
=  \frac{2 \left| P^{x}_{ n n'} ({\bm k})
 \pm i P^{y}_{ n n'} ({\bm k}) \right|^2}{ m_e \hbar \omega_{ n n'}({\bm k}) } 
 \label{oscstr}
\end{eqnarray}
where $P^{x/y}_{ n n' } = m_e \langle u_{n} \mid  \hat{\upsilon}_{x/y} \mid u_{n'} \rangle$ and $\hbar \omega_{ n n'} 
= E_{ n }( {\bm k} ) - E_{ n' } ({\bm k})$ is the energy difference between
the $u_n$ and $u_{n'}$ states, and we define the degree of circular polarization for each $k$-point to be~\cite{yaowang2007}
\begin{eqnarray}
\eta_{n n'} ({\bm k}) = \frac{ \left| P_{n n'}^{+} ({\bm k}) \right|^2  - \left| P_{n n'}^{-} ({\bm k}) \right|^2 }{  \left| P_{n n'}^{+}({\bm k})  \right|^2 + \left| P_{n n'}^{-}({\bm k})  \right|^2 }.
\end{eqnarray}
For sake of simplicity the present analysis focuses on the strongest features of interband transitions in the dipole 
approximation and neglects the cross $\sigma_{xy}$ terms responsible for the optical activity due to twist angle 
dependent phase difference between top and bottom layers~\cite{breydichroism}
and the higher order terms proportional to the magnetic fields ~\cite{gomezsantos2018}.
Studies about the selection rules for the promotion of the excitons~\cite{xiaodi_exciton,tcao}
and discussions related with the moire excitons~\cite{elaine2019,xuxiaodong2019,wangfengexciton2019}
for twisted gapped Dirac bilayer systems will be discussed elsewhere. 
\begin{figure}[htb!]
	\begin{center}
		\includegraphics[width=8cm]{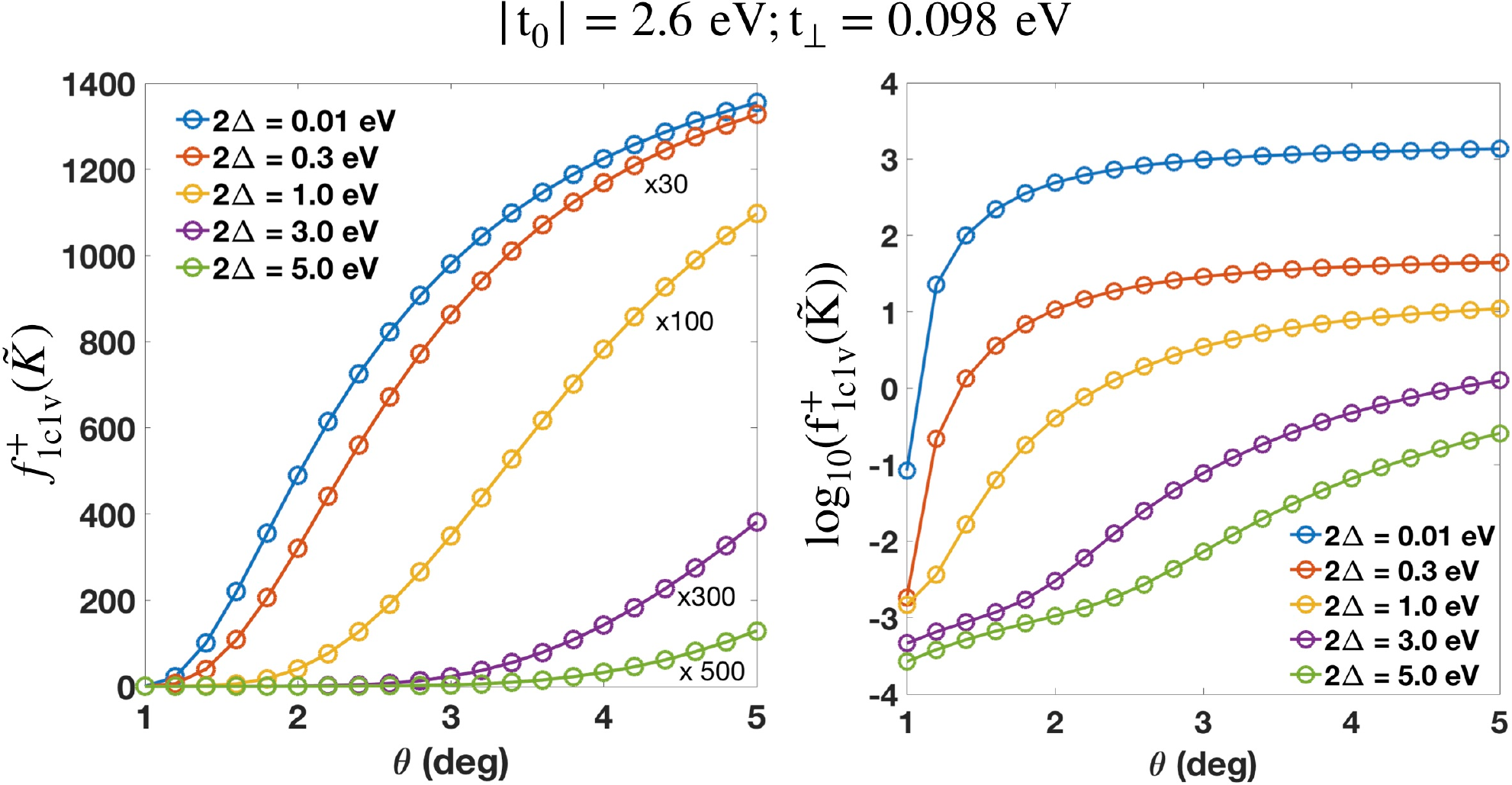}
	\end{center}
	\caption{(Color online) 
	Oscillator strength corresponding to $f_{\rm 1c 1v}^{+}$ transitions between the highest valence and lowest 
	conduction bands evaluated at the mini valley point $\tilde{K}$
	where maxima values are expected in gapped Dirac materials.
	The suppression of the oscillator strength for smaller twist angles and intralayer band gap $2\Delta$
	illustrates the reduction of the quasiparticle velocity associated with the flattening of the bands.
	In the left panel we show that the suppression of the oscillator strength is faster than the growth rate of $2 \Delta$
	for a fixed value of $\theta$. 
	In the right panel we show more clearly the steeper increase of the oscillator strength with twist angle in the small angle regime.
	}
	\label{fevolution}
\end{figure}

In Fig.~\ref{interband} we show a schematic illustration of the band edges near each macrovalley $K$ and $K'$ for twisted bilayers 
near $0^{\circ}$ and 60$^{\circ}$ alignments where we expect different optical responses because the band edges near 
$\tilde{K}$ and $\tilde{K}'$ minivalleys within a macrovalley could have the same or opposite mass signs. 
Circular dichroism is present near $0^{\circ}$ alignment due to the alignment of the mass signs in each macrovalley, 
while near $60^{\circ}$ alignment the valley contrasting circular dichroism is canceled although not completely 
if we take into account the phases acquired due to the rotation of the layers~\cite{jiwoongpark2016,gomezsantos2018} 
that should also be accounted for in our twisted gapped Dirac systems.
We can see that this cancellation can also becomes nonzero when the minivalleys are polarized by the simultaneous presence 
of an external electric field that polarizes layer and carrier doping that shifts the Fermi level.

In Fig.~\ref{interband2} we illustrate how the interband optical transition oscillator strengths 
are progressively modified in twisted systems going from 
two weakly coupled gapped Dirac cones for large twist angles of $\sim 3^{\circ}$ to strongly coupled bands leading to flat band systems near $\sim 1^{\circ}$.
The separate representation of $\left| P^+_{nn'} ({\bm k}) \right|^2$ and $\left| P^-_{nn'} ({\bm k}) \right|^2$
allows to distinguish the sensitivity to opposite circular polarization of the states at different $k$-points in the mBZ.
In the limit of large twist angles the distribution of the oscillator strengths in the mBZ are in qualitative agreement with the expected interband transitions 
obtained juxtaposing two gapped Dirac cones next to each other separated by $K_{\theta}$ in momentum space and coupled through the interlayer tunneling parameters $\omega_i$. 
For large enough twist angles and small primary gaps the interband transition oscillator strengths clearly peak around the 
$\tilde{K}$ and $\tilde{K}'$ points in keeping with the optical properties of a single gapped Dirac Hamiltonian~\cite{xiaodi}. 
As the twist angle is reduced and the interlayer coupling strength is enhanced we observe 
a progressive flattening of the bands and reduction of quasiparticle velocities 
that in turn results in a reduction in the oscillator strength in Eq.~(\ref{oscstr}).  
Its evolution at the mini Dirac points represented in Fig.~\ref{fevolution}
clearly shows a superlinear suppression of the oscillator strengths as a function of the band gap
and that its evolution is faster for smaller than larger twist angles.
However, the almost divergent increase in the density of states $D(E) \propto \theta^{2-m}$ in Eq.~(\ref{dosevol})
due to squeezing bandwidth should increase the absorption rate as the bands become flatter until the flatness is limited
by the broadening width due to disorder or temperature. 
This band flattening leads to smoother spread out and broad distribution of the Berry curvatures near the band edges in the mBZ. 
However, even in the limit of very narrow bandwidths near $\sim 1^{\circ}$ the oscillator strength 
for circularly polarized light remains predominantly centered around the minivalleys $\tilde{K}$ and $\tilde{K}'$ in the mBZ
resulting in a polarization function $\eta({\bm k})$ close to unity near the minivalleys. 
Due to the almost complete flattening of the bands
we notice that an external electric field that shifts the position of the band edges near the minivalley points 
can introduce distortions to the band structure that are significant enough
to modify the oscillator strengths.

\section{summary and discussion}
\label{summarysec}
Recent research on twisted van der Waals materials 
is rapidly expanding beyond twisted bilayer graphene (TBG) to include twisted layered materials with intrinsic gaps.
In this work we have identified and explained the practical advantages of twisted gapped Dirac materials over TBG for the formation of narrow bandwidth flat bands. 
%
We studied the conditions for the generation of narrow bandwidth flat bands as a function of twist angle based on the extended  
Bistritzer-MacDonald model of twisted bilayer graphene with a finite mass term in each one of the layers, allowing for intralayer moire patterns, 
and using up to three different interlayer tunneling parameters $\omega_i$ with $i=1,2,3$ for a more precise description of the interlayer coupling. 
%
%
%
%
%
We have identified the evolution of the low energy bandwidth ($W$) and found its dependence as a function of 
the band gap ($2 \Delta$), twist angles ($\theta$), the Fermi velocity ($\upsilon_{\rm F}$) and interlayer coupling ($\omega_i$). 
The fitting equation Eq.~(\ref{eq1}) is expected to be valid in the parameter range where several realistic 2D material combinations lie
including gapped graphene due to alignment with hexagonal boron nitride (BN), transition metal dichalcogenides (TMDC), silicon carbide (SiC),
and our analysis should be valid when the band edges near the $K$ point can be described with gapped Dirac Hamiltonians.
%
%

One of the main conclusions we draw is that a finite gap in the constituent layers 
of the twisted gapped Dirac materials makes the generation of flat bands simpler than in TBG because already for 
band gaps of $\sim$250~meV the band flattening in twisted bilayers
happens for a continuous range of small twist angles without requiring specific magic angles.
Moreover, the larger the gaps the greater the suppression of the band width allowing to achieve narrower bands for 
similar twist angles, a fact that should facilitate achieving stronger effective Coulomb interactions $U \propto \theta$ that scales with twist angle.
In moderately gapped TMDC materials or large band gap hBN materials we find that the bandwidths can remain below $\sim10$~meV 
even for twist angles as large as $\sim3^{\circ}$.   
%
%
Stronger interlayer coupling parameters $\omega_i$ also allows to achieve narrow bandwidths for larger twist angles.
In the example case of twisted SiC bilayers perturbed by relatively strong intralayer moire patterns and 
interlayer coupling the interplay of three unequal $\omega_i$ 
gave rise to valence bands with bandwidths on the order of $\sim$20~meV even for twist angles as large as $\sim$7$^{\circ}$,
which implies a seven fold enhancement of the Coulomb interaction strength with respect to magic angle TBG based on the scale
of the moire pattern periods.
%
%
Our conclusions based on numerical calculations are complemented by the analytical solutions of 
the band eigenvalues at the symmetry points of the mBZ that provides estimates for the bandwidth as a function of 
the different system parameters such as twist angle, interlayer coupling and band gap. 
%
%

%
The topological nature of the nearly flat bands in twisted gapped Dirac materials have been studied by 
calculating the phase diagrams for the valley Chern numbers associated to the first four conduction and valence bands. 
In particular, it was shown that the interlayer tunneling terms $\omega_i$ have to be different from each other for a Chern 
band to emerge in the limit of small intralayer band gaps like in systems of graphene on hexagonal boron nitride.
The increase of twist angle $\theta$ in general helps to expand the phase space for the low energy valence and conduction 
bands to acquire a finite valley Chern number although this can effectively weaken the influence of the interlayer moire patterns 
and the magnitude of the isolation secondary gaps.
For larger gap systems like in semiconducting TMDC or in the limit of large gap systems like hexagonal boron nitride, 
the lowest energy levels of a twisted bilayer remain topologically trivial but the higher bands can remain topological for large enough twist angles. 
These higher energy bands should be accessible either by gating techniques in devices in the small twist angle limit when 
the electron density per band is small, or through optical measurements. 
%
%
%
%
%
%
The valley contrasting circular dichroism in twisted gapped Dirac materials inherits the properties of 
single gapped Dirac layers whose band edges locate at the minivalley $\tilde{K}$ and $\tilde{K}'$ points of the moire Brillouin zone
around which maxima in the optical transition oscillator strength for circularly polarized light and Berry curvature are often found. 
This behavior was found to persist even in the limit where multiple atomic level like nearly flat bands
were present where the traces of the original gapped Dirac cone band edges could not be clearly identified. 
Qualitatively distinct optical response to circularly polarized light is expected between 
twisted bilayer systems near $0^{\circ}$ and 60$^{\circ}$ (or equivalently 180$^{\circ}$)
alignment where the phase winding and Berry curvature values of the bands in each minivalley could be aligned to point 
in the same or opposite directions respectively. 

In summary, our analysis suggests optimistic prospects of finding isolated flat bands in a variety of twisted gapped Dirac materials
other than twisted bilayer graphene, in particular when the original bandwidth of the building block materials 
are narrow to begin with, or when a strong interlayer interaction allows to access nearly flat bands for a larger range of twist angles.

\begin{acknowledgments}
This work was supported by the Samsung Science and Technology Foundation under project no. SSTF-BA1802-06 for J. S., 
and from the Korean NRF grant number NRF-2016R1A2B4010105 for S. J.
The 2017 Research Fund of the University of Seoul is acknowledged for J. J.
Financial support for J. S. has also been granted by the National Natural Science Foundation of China (Grant No. 11604166), 
Zhejiang Provincial Natural Science Foundation of China (Grant No. LY19A040003) and K. C. Wong Magna Fund in Ningbo University.
This work was partly performed at the Aspen Center for Physics, which is supported by National Science Foundation grant PHY-1607611.
\end{acknowledgments}


\begin{appendix}
\section{Hamiltonian parameters fitting procedure from DFT calculations}
\label{fitting}

Here we outline the methodology followed to obtain the Dirac Hamiltonian model parameters for materials involving SiC that
were calculated from LDA-DFT calculations with the Perdew-Zunger parametrization~\cite{perdewzunger}. 
For SiC bilayers we started from the full tight-binding (FTB) model Hamiltonian $H_{mono}$ for monolayer SiC using the hopping parameters extracted from 
the maximally localized Wannier functions by means of a two by two Hamiltonian~\cite{maximallylocalizedwannier} by means of a two by two Hamiltonian
	\begin{eqnarray}
		H_{\rm mono} = \begin{bmatrix}H_{Si-Si} & H_{Si-C}\\ H_{C-Si}& H_{C-C} \end{bmatrix}.
	\end{eqnarray}
This monolayer Hamiltonian is then used to construct the bilayer SiC Hamiltonian through 	
	\begin{eqnarray}
	H_{\rm bilayer} = \begin{bmatrix}H_{\rm mono} & T^0\\ T^{0*}& H_{\rm mono} \end{bmatrix}
	\end{eqnarray}
where, is the  FTB model Hamiltonian of monolayer SiC, and T$^0$ is the interlayer tunneling matrix as given in Eq.~(\ref{Eq:Tmatrix}).
We have determined the tunneling matrix elements $\omega_i$ that define $T_0$ by assuming that for large gap systems
$\omega_1$, $\omega_3$ determines the splitting in conduction, valence bands respectively, whereas $\omega_2$ controls the interaction between conduction and valence bands.
When fitting the tunneling matrix elements to reproduce the DFT-bands near the $K$-point for different standard stacking structures (AA, AB and BA)
we followed the criteria that (a) the low energy bands should be parabolic at K-point so as to match with the gapped Dirac cone shaped DFT bands, (b) for all three standard SiC bilayer
staking, at least four low energy bands (2-conduction,2-valence) should reproduce the DFT bands near K-point, and (c) the optimum $\omega_i$ parameters set should be unique and reproducible.
With these criteria in mind the optimum tunneling parameters for AA-stacked bilayer SiC, are $\omega_1$ = 0.485 eV, $\omega_2$ = $\omega_2^*$ = 0, and $\omega_3$ = 0.19 eV,
while for AB-stacking, $\omega_1$ = $\omega_2$ = $\omega_3$ = 0 and $\omega_2^*$ = 1.24 eV, and for BA-stacking $\omega_1$ = $\omega_2^*$ = $\omega_3$ = 0 and $\omega_2$ = 1.24 eV.
The tunneling matrix elements obtained from above procedure are then averaged for the three standard stacking structures
and they are $\omega_1$ = 0.165 eV, $\omega_2$= $\omega_2^*$ = 0.413 eV and $\omega_3$ = 0.063 eV as listed in Table I.

\section{Analytical wave functions and comparison with numerical calculations}
\label{analytical}
In this appendix we provide further details on the analytical solutions of the wave functions evaluated at the symmetry points. 
As we have discussed in the main text, $\{\eta_1, \eta_2\}$ only have three different combinations, which are
$\phi_1= \{0, 2\pi/3\}$, $\phi_2= \{2\pi/3, -2\pi/3\}$, and  $\phi_3=\{-2\pi/3, 0\}$. For each $\phi_i$ ($i=$1,2,3), the eigenenergy problem is then changed to solving a quartic equation 
\begin{eqnarray}
x^4 + \alpha x^2 + \beta x + \gamma =0,
\label{Eq:quartic}
\end{eqnarray}
which has a general formula for roots, but the roots are in quite complicated forms. 
For three different $\phi$s, the coefficient in Eq.~(\ref{Eq:quartic}) have different values and we discuss the value and details of calculations of $E_{\phi_i}$s below.
For $\phi_1$, the coefficients are 
\begin{equation}
\begin{aligned}
\alpha &= -2(\Delta^2 + \rho_\theta^2 + 2.5\omega_1^2+\omega_2^2), \\
\beta &= -6\Delta \omega_1^2,\\
\gamma &= (\Delta^2 + \rho_\theta^2)^2 -\Delta^2(5\omega_1^2-2\omega_2^2)-\rho_\theta^2(4\omega_1^2+2\omega_2^2)\\ 
&+ (2\omega_1^2+\omega_2^2)^2. 
\end{aligned}\label{Eq:coeff_phi1}
\end{equation}
If $\beta=0$, Eq.~\ref{Eq:quartic} becomes a quadratic equation and one of the roots is 
\begin{widetext}
	\begin{equation}
	\begin{aligned}
	 E_{\phi_1}^0(\tilde{\Gamma}) =\sqrt{\Delta^2 +\rho_\theta^2+2.5\omega_1^2 + \omega_2^2 -\sqrt{10\Delta^2\omega_1^2 + 9\rho_\theta^2\omega_1^2 +9/4\omega_1^4 + 4 \rho_\theta^2\omega_2^2 + \omega_1^2\omega_2^2}  }. 
	\end{aligned}
	\end{equation}
\end{widetext}
Due to the ignorance of the $\beta=-6\Delta\omega_1^2$ term, $E_{\phi_1}^0(\tilde{\Gamma})$ is a good approximation only for small $\Delta$ and $\omega_1$. The energy correction $\delta E_{\phi_1}$ can be obtained by 
substituting $x=E_{\phi_1}^0(\tilde{\Gamma}) - \delta E_{\phi_1}$ to Eq.~\ref{Eq:quartic} with the coefficients given by Eq.~\ref{Eq:coeff_phi1}, and expanding to the first order of $\delta E_{\phi_1}$. After a straight forward calculation, one can obtain 
\begin{equation}
\begin{aligned}
\delta E_{\phi_1} = \frac{3\Delta\omega_1^2}{2\sqrt{10\Delta^2\omega_1^2 + 9\rho_\theta^2\omega_1^2 +9/4\omega_1^4 + 4 \rho_\theta^2\omega_2^2 + \omega_1^2\omega_2^2}}, 
\end{aligned}
\end{equation}
and hence 
\begin{equation}
\begin{aligned}
E_{\phi_1}(\tilde{\Gamma})  \approx E_{\phi_1}^0(\tilde{\Gamma}) -\delta E_{\phi_1}. 
\end{aligned}
\end{equation}
For $\phi_3$ case, $\alpha$ and $\gamma$ values for are the same as those in the $\phi_1$ case as given in Eq.~\ref{Eq:coeff_phi1}, the only difference is that $\beta=6\Delta \omega_1^2$, which is with opposite sign in comparison with Eq.~\ref{Eq:coeff_phi1}. Hence $E_{\phi_3}(\tilde{\Gamma})$ has a similar form with $E_{\phi_1}(\tilde{\Gamma})$ and is given by 
\begin{equation}
\begin{aligned}
E_{\phi_3}(\tilde{\Gamma})  \approx E_{\phi_1}^0(\tilde{\Gamma}) +\delta E_{\phi_1}. 
\end{aligned}
\end{equation} 

\begin{figure}[htb!]
	\begin{center}
		\includegraphics[width=8cm]{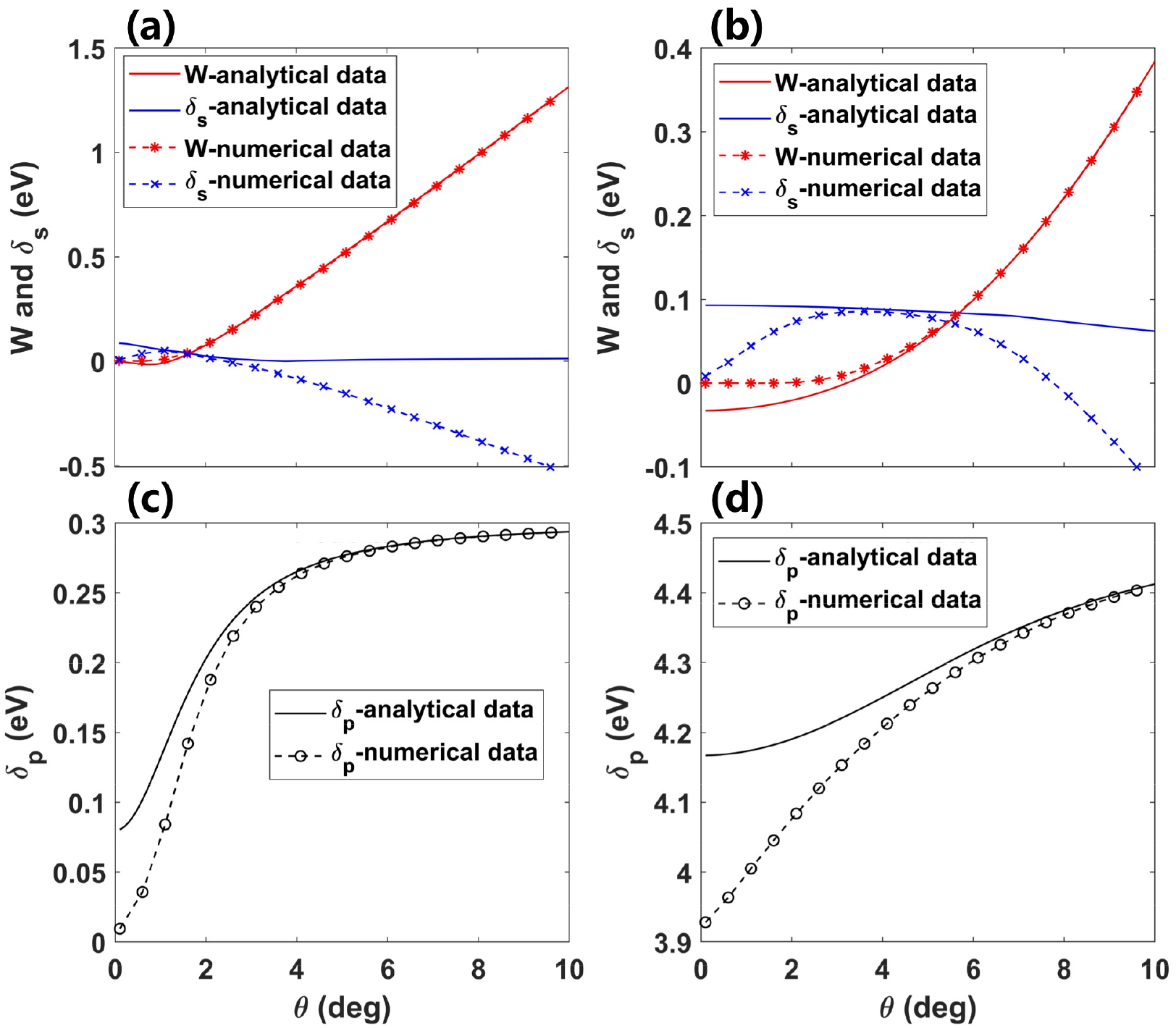}
	\end{center}
	\caption{(Color online) Comparison of the analytical and the numerical results for $\omega_1=\omega_2=\omega_3=0.098$eV and $t_0=-2.6$eV. 
	The left column shows the results for $2\Delta=0.3$eV corresponds to a small gap case, and the right column is for a large value $2\Delta=4.5$eV 
	which is close to the gap size in BN/BN. 
	We find a satisfactory agreement between both results especially in the regime of large twist angles where the analytical results are expected to work better. 
	The difference of $\delta_s$ between the numerical and analytical data defined as $\delta_s = E_{2}(\Gamma) - E_{1}(\Gamma)$ 
	for large twist angles is because the second conduction (valence) 
	band minimum (maximum) starts to happen away from the $\Gamma$ point rather than the due to lack of accuracy in the analytical expressions. 
	}	
	\label{supp_w}
\end{figure}
The coefficients for $\phi_2$ case are 
\begin{equation}
\begin{aligned}
\alpha &= -2(\Delta^2 + \rho_\theta^2 + \omega_1^2+4\omega_2^2), \\
\beta &= 0,\\
\gamma &= (\Delta^2 + \rho_\theta^2)^2 -\Delta^2(2\omega_1^2-8\omega_2^2)-\rho_\theta^2(-2\omega_1^2+8\omega_2^2) \\
&+ (\omega_1^2-4\omega_2^2)^2. 
\end{aligned}\label{Eq:coeff_phi2}
\end{equation}
The energy corresponds to $\phi_2$ at $\tilde{\Gamma}$ point has a simpler form 
\begin{widetext}
	\begin{eqnarray}
	E_{\phi_2}(\tilde{\Gamma})=\sqrt{\Delta^2 + \rho_\theta^2 + \omega_1^2 + 4\omega_2^2 -2\sqrt{\Delta^2\omega_1^2+4\rho_\theta^2\omega_2^2+4\omega_1^2\omega_2^2}}
	\end{eqnarray}
\end{widetext}

When $\omega_1=\omega_3$, the gapped Dirac system preserves the electron-hole symmetry. We briefly discuss the energy eigenvalues the related $\phi_i$ phases for the three valence bands close to the Fermi level. 
As we have introduced in Section. IV. B, for each $\phi$ one may obtain 4 roots, which are 4 eigenenergies of the 12 by 12 Hamiltonian. The three lowest conduction band energies are given by $E_{\phi_i}(\tilde{\Gamma})$ ($i=$1,2,3), where $E_{\phi_i}(\tilde{\Gamma})$ is one of the 4 roots when the vector phases are given by $\phi_i$. Among the other three roots for each $\phi_i$, there exists one root which corresponds to one of the three valence bands close to the Fermi level. The valence bands energies are 
\begin{widetext}
 \begin{equation}
 \begin{aligned}
 E_{\phi_1 / \phi_3}^v(\tilde{\Gamma}) & \approx -\sqrt{\Delta^2 +\rho_\theta^2+2.5\omega_1^2 + \omega_2^2 -\sqrt{10\Delta^2\omega_1^2 + 9\rho_\theta^2\omega_1^2 +9/4\omega_1^4 + 4 \rho_\theta^2\omega_2^2 + \omega_1^2\omega_2^2}  }\\ 
 & \mp \frac{3\Delta\omega_1^2}{2\sqrt{10\Delta^2\omega_1^2 + 9\rho_\theta^2\omega_1^2 +9/4\omega_1^4 + 4 \rho_\theta^2\omega_2^2 + \omega_1^2\omega_2^2}},\\ 
 E_{\phi_2}^v(\tilde{\Gamma})&=-\sqrt{\Delta^2 + \rho_\theta^2 + \omega_1^2 + 4\omega_2^2 -2\sqrt{\Delta^2\omega_1^2+4\rho_\theta^2\omega_2^2+4\omega_1^2\omega_2^2}}.  
 \end{aligned}
 \end{equation}
\end{widetext}
One can find that the valence band energies at the $\tilde{\Gamma}$ point are related to the conduction band energies as $E_{\phi_1}^v(\tilde{\Gamma})=-E_{\phi_3}(\tilde{\Gamma})$, $E_{\phi_2}^v(\tilde{\Gamma})=-E_{\phi_2}(\tilde{\Gamma})$ and 
$E_{\phi_3}^v(\tilde{\Gamma})=-E_{\phi_1}(\tilde{\Gamma})$.

%
\begin{figure}[h!]
	\begin{center}
		\includegraphics[width=8cm]{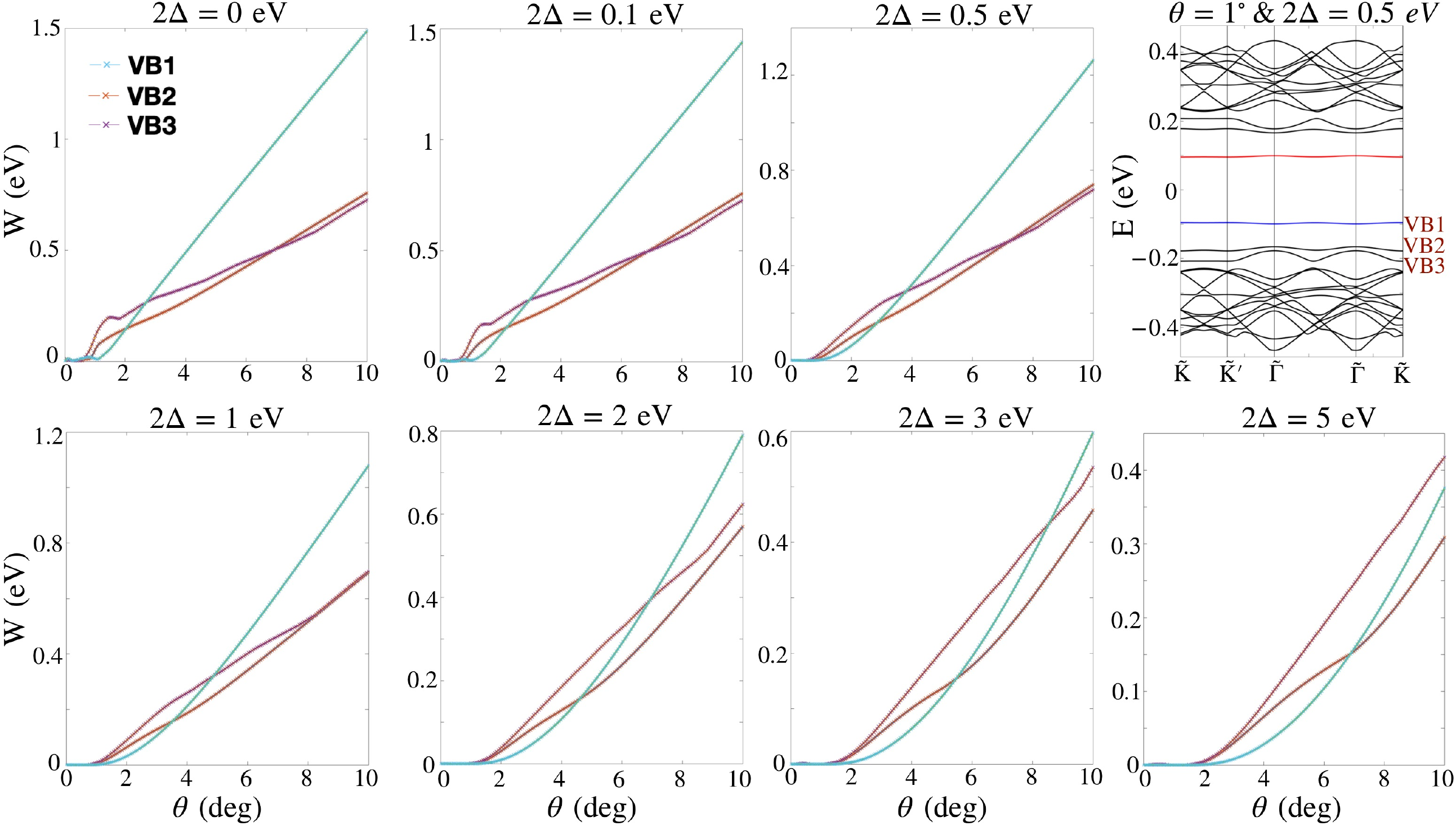}
	\end{center}
	\caption{(Color online) 
	Evolution of the bandwidth corresponding to the first three low energy bands for different band gaps $2 \Delta$. 
	We observe a clear decrease in the bandwidths of the low lying energy bands when the band gaps become larger. 
	The bandwidth narrowing is more effective for the lowest energy conduction or valence band represented with 
	a green line extending for larger $\theta$ values as the band gap $2\Delta$ increases.
	}
	\label{higherbands}
\end{figure}
\section{Bandwidth evolution in higher energy bands}
\label{hibands}
As noted in the main text the bandwidth of the moire energy bands are reduced when
the twist angles become sufficiently small or when the intralayer gaps are increased. 
The narrowing of the bandwidth in the three low lying conduction or valence bands for increasing 
$2\Delta$ is illustrated in Fig.~\ref{higherbands} and is also reflected in the 
progressive reduction of the y-axis scale. 
For the Hamiltonian approximation where we have used with equal $\omega_i$ and neglected the twisting phases we have electron-hole symmetry
that results in equal bandwidths for the valence and conduction bands. 
The bandwidth compression happens most effectively for the lowest energy valence and conduction bands as the band gap $2\Delta$ is increased.

\section{Mass dependent Chern number phase diagram for the valence bands}
\label{massdep}
\begin{figure*}[h]
	\begin{center}
		\includegraphics[width=16cm]{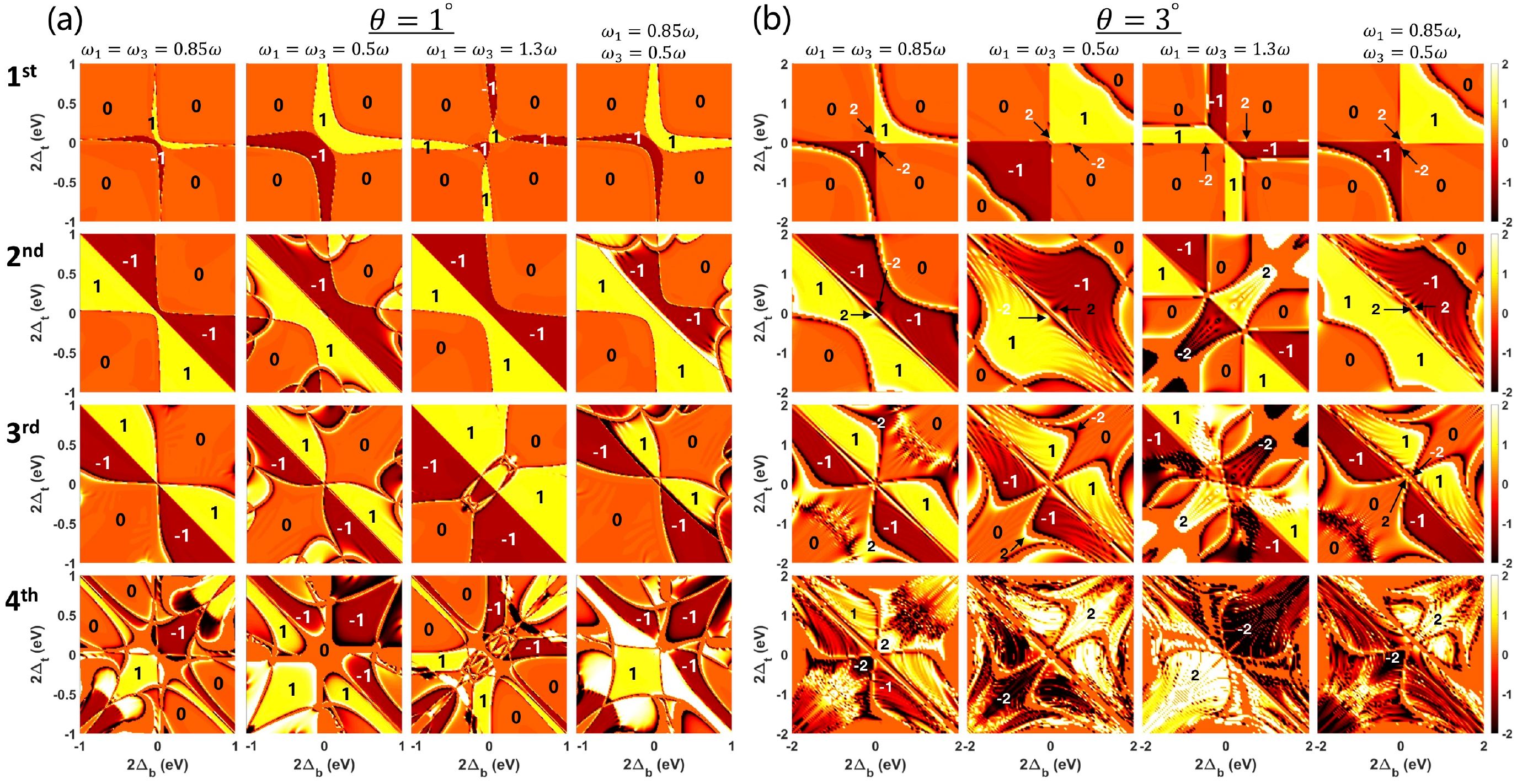}
	\end{center}
	\caption{(Color online)
		Valley Chern number phase diagram of massive twisted bilayer graphene as a function of the mass terms $\Delta_{t}$ and $\Delta_{b}$ for the top and bottom layers for the first four valence bands for a variety of massive twisted bilayer graphene systems. This figure resembles the phase diagram presented in the main text for the conduction bands in Fig.~\ref{chern_angle}.
	}	
	\label{figure11_valence}
\end{figure*}

The Chern number phase diagram for the valence bands shown in Fig.~\ref{figure11_valence} closely resemble those of the conduction bands represented in 
Fig.~\ref{figure11} in virtue of the overall electron-hole symmetry in our model when the twist angle dependent phases are not included.
The deviations in the electron-hole symmetry that distinguish the results of the valence and conduction bands 
stem from the use of unequal interlayer tunneling values $\omega_i$.

\label{vbandsphasediag}

\end{appendix}

\newpage

\end{document}